\documentclass[10pt,conference,letterpaper]{IEEEtran}
\usepackage{times,amsmath,epsfig}
\usepackage{latexsym}
\usepackage{multirow}
\usepackage{epsfig}
\usepackage{amsfonts}
\usepackage[tight]{subfigure}
\usepackage{algorithm,algorithmic}
\newcommand{\comment}[1]{}

\newsavebox{\boxone}
\newsavebox{\boxtwo}
\newsavebox{\boxthree}

\newlength{\narrow}
\newlength{\cnarrow}
\newcommand{\topline}{
  \hrule
  \vskip .5\baselineskip}
\newcommand{\bottomline}{
  \vskip 2pt
  \hrule}

\newcommand{\chbox}[2]{
  \hbox to #1{\hss\vtop{#2}\hss}}

\newcommand{\nchbox}[1]{
  \chbox{\narrow}{#1}}

\newcommand{\cnchbox}[1]{
  \chbox{\cnarrow}{#1}}

\newcommand{\fcode}[1]{
  
  \chbox{\textwidth}{\tgrind\input{#1}}}

\newcommand{\ncode}[1]{
  
  \chbox{\narrow}{\tgrind\input{#1}}}

\def\nfig#1#2#3{
  \vtop{\nchbox{#1}
  \hbox to\narrow{\parbox{\narrow}{\caption{#2}\label{#3}}}}}

\newcommand{\cncode}[1]{
  \chbox{\cnarrow}{\tgrind\input{#1}}}

\def\codefiggen[#1]#2#3#4#5#6{
  \begin{figure}[#1]
  #5
  \fcode{#2}
  \center\parbox{.9\textwidth}{\caption{#3}\label{#4}}
  #6
  \end{figure}}

\def\codefig[#1]#2#3#4{
  \codefiggen[#1]{#2}{#3}{#4}{}{}}

\def\codefigline[#1]#2#3#4{
  \codefiggen[#1]{#2}{#3}{#4}{\topline}{\bottomline}}

\def\doublefiggen[#1]#2#3#4#5#6#7#8#9{
  \begin{figure}[#1]
  #8
  \hbox to \textwidth{
  \nfig{#2}{#3}{#4}
  \hfil
  \nfig{#5}{#6}{#7}}
  #9
  \end{figure}}

\def\doublefig[#1]#2#3#4#5#6#7{
  \doublefiggen[#1]{#2}{#3}{#4}{#5}{#6}{#7}{}{}}

\def\doublefigline[#1]#2#3#4#5#6#7{
  \doublefiggen[#1]{#2}{#3}{#4}{#5}{#6}{#7}{\topline}{\bottomline}}

\def\doublecodefig[#1]#2#3#4#5#6#7{
  \doublefig[#1]{\ncode{#2}}{#3}{#4}{\ncode{#5}}{#6}{#7}}

\def\doublecodefigline[#1]#2#3#4#5#6#7{
  \doublefigline[#1]{\ncode{#2}}{#3}{#4}{\ncode{#5}}{#6}{#7}}

\newcommand{\codepair}[4]{\vbox{
  \hbox{\ncode{#1} \hfil \ncode{#3}}
  \vskip .3\baselineskip plus .3\baselineskip
  \hbox{\hbox to\narrow{#2\hfil} \hfil \hbox to\narrow{#4\hfil}}}}

\def\codepairfig[#1]#2#3#4#5#6#7{
  \begin{figure}[#1]
  \codepair{#2}{#3}{#4}{#5}
  \center\parbox{.9\textwidth}{\caption{#6}}
  \label{#7}
  \end{figure}}

\def\cncodepairfiggen[#1]#2#3#4#5#6#7{
  \begin{figure}[#1]
  #6
  \hbox{\cncode{#2}\hfil\cncode{#3}}
  \center\parbox{.9\columnwidth}{\caption{#4}\label{#5}}
  #7
  \end{figure}}

\def\cncodepairfig[#1]#2#3#4#5{
  \cncodepairfiggen[#1]{#2}{#3}{#4}{#5}{}{}}

\def\cncodepairfigline[#1]#2#3#4#5{
  \cncodepairfiggen[#1]{#2}{#3}{#4}{#5}{\topline}{\bottomline}}

\def\doublefigOnecap*[#1]#2#3#4#5{
  \begin{figure*}[#1]
  \hbox to \textwidth{
  \nchbox{#2}
  \hfil
  \nchbox{#3}}
  \caption{#4}
  \label{#5}
  \end{figure*}}

\def\doublefigOnecap[#1]#2#3#4#5{
  \begin{figure}[#1]
  \topline
  \hbox to \columnwidth{
  \cnchbox{#2}
  \hfil
  \cnchbox{#3}}
  \caption{#4}
  \label{#5}
  \bottomline
  \end{figure}}

\def\PSfig[#1]#2#3#4{
 \begin{figure}
 \centerline{\psfig{file=#2,width=\columnwidth}}
 \caption{{#3}}
 \label{#4}
 \end{figure}}

\def\PSfiglines[#1]#2#3#4{
 \begin{figure}[#1]
 \topline
 \centerline{\psfig{file=#2,width=\columnwidth}}
 \caption{{#3}}
 \label{#4}
 \bottomline
 \end{figure}}

\def\PSfiglinesht[#1]#2#3#4#5{
 \begin{figure}[#1]
 \topline
 \centerline{\psfig{file=#2,height=#3}}
 \caption{{#4}}
 \label{#5}
 \bottomline
 \end{figure}}

\def\doublePSfig[#1]#2#3#4#5#6{
  \doublefigOnecap[#1]
    {\cnchbox{\psfig{file=#2,height=#4}}}
    {\cnchbox{\psfig{file=#3,height=#4}}}
    {#5}
    {#6}}

\newlength{\boxwidth}
\newcommand{\bproof}{{\bf Proof Sketch:}}
\newcommand{\eproof}{\mbox{$\Box$}}

\def\tabdoublecode#1#2#3#4{
 \begin{figure*}[t]
 \topline\vs{-.4}
 \hbox to \columnwidth{
 \vtop{\small
 \begin{tabbing}
 #1
 \end{tabbing}}
 \hfil
 \hfil
 \hfil
 \vtop{\small
 \begin{tabbing}
 #2
 \end{tabbing}}
 }
 \caption{#3\label{#4}}
 \bottomline
 \end{figure*}
}
\def\tabtriplecode#1#2#3#4#5{
 \begin{figure}
 \topline\vs{-.4}
 \hbox to \columnwidth{
 \vtop{\small
 \begin{tabbing}
 #1
 \end{tabbing}}
 \hfil
 \hfil
 \hfil
 \vtop{\small
 \begin{tabbing}
 #2
 \end{tabbing}}
 \hfil
 \hfil
 \hfil
 \vtop{\small
 \begin{tabbing}
 #3
 \end{tabbing}}
 }
 \caption{#4\label{#5}}
 \bottomline
 \end{figure}
}

\newtheorem{lemma}{Lemma}
\newcommand{\blemma}{\begin{lemma}}
\newcommand{\elemma}{\end{lemma}}

\newtheorem{thm}{Theorem}
\newcommand{\bthm}{\begin{thm}}
\newcommand{\ethm}{\end{thm}}

\newtheorem{defin}{Definition}
\newcommand{\bdefin}{\begin{defin}}
\newcommand{\edefin}{\end{defin}}

\newtheorem{observation}{Observation}
\newcommand{\bobserv}{\begin{observation}}
\newcommand{\eobserv}{\end{observation}}

\newcommand{\vs}[1]{\vspace{#1cm}}
\newcommand{\be}{\begin{equation}}
\newcommand{\ee}{\end{equation}}

\newcommand{\bdesc}{\begin{description}}
\newcommand{\edesc}{\end{description}}
\newcommand{\benum}{\begin{enumerate}}
\newcommand{\eenum}{\end{enumerate}}
\newcommand{\bitem}{\begin{itemize}}
\newcommand{\eitem}{\end{itemize}}
\newcommand{\bcenter}{\begin{center}}
\newcommand{\ecenter}{\end{center}}
\newcommand{\btabular}{\begin{tabular}}
\newcommand{\etabular}{\end{tabular}}
\newcommand{\beqnarr}{
 \begin{eqnarray}}
\newcommand{\eeqnarr}{\end{eqnarray}}

\newtheorem{mydef}{Definition}

\title{Large Scale Real-time Ridesharing with Service Guarantee on Road Networks}

\author{%
{ Yan Huang{\small $~^{\#1}$}, Ruoming Jin{\small $~^{*2}$}, Favyen Bastani{\small $~^{\$3}$}, Xiaoyang Sean Wang {\small $~^{+4}$} }%
\vspace{1.6mm}\\
\fontsize{10}{10}\selectfont\itshape
$~^{\#}$Computer Science and Engineering, University of North Texas\\
\fontsize{9}{9}\selectfont\ttfamily\upshape
$~^{1}$huangyan@unt.edu
\vspace{1.2mm}\\
\fontsize{10}{10}\selectfont\rmfamily\itshape
$~^{*}$ Computer Science, Kent State University\\
\fontsize{9}{9}\selectfont\ttfamily\upshape
$~^{2}$jin@cs.kent.edu 
\vspace{1.2mm}\\
\fontsize{10}{10}\selectfont\rmfamily\itshape
$~^{\$}$  Massachusetts Institute of Technology\\
\fontsize{9}{9}\selectfont\ttfamily\upshape
$~^{3}$fbastani@mit.edu%
\vspace{1.2mm}\\
\fontsize{10}{10}\selectfont\rmfamily\itshape
$~^{+}$ Computer Scicence, Fudan University\\
\fontsize{9}{9}\selectfont\ttfamily\upshape
$~^{4}$xywangCS@fudan.edu.cn
}

\begin{document}
\maketitle

\baselineskip=1.0\normalbaselineskip

\begin{abstract}
The mean occupancy rates of personal vehicle trips in the United States is only 1.6 persons per vehicle mile. Urban traffic gridlock is a familiar scene. Ridesharing has the potential to solve many environmental, congestion, and energy problems. In this paper, we introduce the problem of large scale real-time ridesharing with service guarantee on road networks. Servers and trip requests are dynamically matched while waiting time and service time constraints of trips are satisfied. We first propose two basic algorithms: a branch-and-bound algorithm and an integer programing algorithm. However, these algorithm structures do not adapt well to the dynamic nature of the ridesharing problem. Thus, we then propose a kinetic tree algorithm capable of better scheduling dynamic requests and adjusting routes on-the-fly. We perform experiments on a large real taxi dataset from Shanghai. The results show that the kinetic tree algorithm is faster than other algorithms in response time.\end{abstract}

%\category{H.2.8}{Database management}{Database Applications}[graph indexing and querying]

%\terms{Performance}

%\keywords{Graph indexing, Reachability queries, Transitive closure, 3-Hop, 2-Hop, Path-tree} % NOT required for Proceedings

\section{Introduction\label{sec:intro}}
Despite our struggle with energy, pollution, and congestion, many private and public vehicles continue to travel with empty seats. The mean occupancy rate of personal vehicle trips in the United States is only 1.6 persons per vehicle mile \cite{Ghoseiri:2011}. In 1999, if 4\% of drivers  would rideshare it would have offset the increase in congestion in the 68 urban areas completely that year \cite{Billerburg:2011}. Several large cities begin to encourage taxi sharing.

Real-time ridesharing \cite{Tickengo:2011,Stach:2011,Gidofalvi:2008}, enabled by low cost geo-locating devices, smartphones, wireless networks, and social networks, is a service that dynamically arranges ad-hoc shared rides. In a real-time ridesharing with service guarantees on road networks problem (hereafter referred to simply as ridesharing), a set of servers travel over a road network, cruising when not committed to any service and delivering passengers otherwise.  Requests for rides are received over time, each consisting of two points, a {\em source} and a {\em destination}. Each request also specifies two constraints, a {\em waiting time} defining the latest time to be picked up and a {\em service constraint} defining the acceptable extra detour time from the shortest possible trip duration. When a new request is received, it is evaluated immediately by all servers. In order to be assigned the request, a server must satisfy all constraints, both those of the new request and those of requests already assigned to the server. {\em The goal is to schedule requests in real-time and minimize the servers' traveling times to complete all of the committed service while meeting service quality guarantee}.

%These servers can be personal vehicles available for ridesharing or public vehicles such as taxis and buses.

% todo: describe dial-a-ride problem briefly

The traditional dial-a-ride problem \cite{Feuerstein:2001} aims at designing vehicle routes and schedules for small to middle sized trip and vehicle sets, e.g. a few vehicles serving tens of requests.  Large scale private car sharing and real-time on-demand taxi or cab sharing are becoming increasingly popular. Increasing numbers of users use mobile devices or the Internet to request and participate in these ride-sharing services. Tickengo \cite{Tickengo:2011}, founded in 2011, is an open ride system where over 50,000 people participate in ridesharing. Other companies include Avego, PickupPal, Zimride, and Zebigo.  In an urban city like Shanghai, there are approximately 120,000 road intersections, 40,000 taxis, and more than 400,000 taxi trips per day (these numbers are derived from our experimental dataset). Slight change of weather such as light rain will send the city into a gridlock. With the mounting  energy, pollution, and congestion problems in the urban and metropolitan areas that are growing at tremendous rates and already host more than half of the entire human population, trading a small amount of privacy and convenience for energy and cost savings using ridesharing is promising and maybe inevitable. 

However, providing ridesharing service at the urban scale is a non-trivial problem. The core is to devise a real-time matching algorithm that can quickly determine the best vehicle (taxi, cab, bus) to satisfy incoming service requests.  Traditional solutions to the related dial-a-ride problems using branch-and-bound \cite{Colorni:2001} or mixed integer programing \cite{Cordeau:2006} approaches are not designed to deal with these enormous modern situations. Furthermore, most previous solutions focus on scenarios where requests are known ahead of time and servers originate and finish at known depots. The dynamic and en route nature renders many of these algorithms either inapplicable or inefficient. 

In this paper, we focus on developing fast matching algorithms for large scale real-time ridesharing. Our algorithms are applicable to the existing services including taxi services, private vehicle sharing, elevator systems, minibus services, and courier services. We note that there are other important factors which need to be considered for the emerging large scale real-time ridesharing, such as inter-personal (female vs. male), safety, social discomfort, and pricing concerns. Those can be addressed by real-name profiling, reputation, or social network trust building systems \cite{Stach:2011,Ghoseiri:2011} and  beyond the scope of this paper. 

%Our algorithms can also be extended to handle these additional factors, but those are beyond the scope of this paper. 

\comment{
For large-scale ridesharing, devising a real-time matching algorithm is challenging. For example, However, }

\subsection{Problem Definition}
\label{problem}

A road network $G = <V,E,W>$ consists of vertex set $V$ and edge set $E$. Each edge $(u,v) \in E$ ($u,v \in V$) is associated with a weight $W(u,v)$ which indicates the traveling cost along the edge $(u,v)$; this traveling cost can be either a time measure or a distance measure. Assuming driving speeds are available, time measure and distance measure can often be converted from one to another and are used interchangeably.

%todo: $d(s,e)=\min_{p} W(p)$ maybe suggests that it's minimal cost of ALL paths, not just those that link?
%also maybe define something for shortest _path_
Given two nodes $s$ and $e$ in the road network, a path $p$ between them is a vertex sequence $(v_0, v_1, \cdots, v_k)$, where $(v_i,v_{i+1})$ is an edge in $E$, $v_0=s$, and $v_k=e$. The path cost $W(p)=\sum W(v_i,v_{i+1})$ is the sum of each edge cost $W(v_i,v_{i+1})$ along the path. 
The shortest path cost $d(s,e)$ is defined as the minimal cost for paths linking from $s$ to $e$, i.e., $d(s,e)=\min_{p} W(p)$ and the corresponding path with cost $d(s,e)$ is the shortest path from $s$ to $e$.

\bdefin({\bf Trip Request})
A trip $tr=<s,e,w,\epsilon>$ with respect to a road network $G = <V,E,W>$ is defined by a source $s \in V$, a destination $e \in V$, a maximal waiting time $w$ (defines the maximum time allowed between making the request and receiving the service), and a service constraint $\epsilon$ for the extra detour time in a trip (bounding the overall distance from $s$ to $e$ by $(1+\epsilon)d(s,e)$). 
\edefin

We consider a unified waiting time $w$ and service constraint $\epsilon$ for all requests, which can be specified by the service provider. However, our proposed algorithms can be easily generalized to individualized waiting time and service constraints. We further assume that $G$ is static over time (i.e., we do not consider different path costs at different times of the day), but the algorithms we present can handle the case where $G$ changes under a predetermined course, and can be extended to the case where $G$ changes unpredictable (for example, to simulate dynamic traffic conditions).

To deal with real-time ride sharing, for each trip $tr_i=<s_i,e_i,w,\epsilon>$ and a given server (e.g., a taxi, cab, or private vehicle), we further introduce $r_i$, the server's location when the request is made. Given this, a general {\em trip schedule} for a server with $m$ trips can be described in a sequence with $3m$ elements, $(x_1,x_2, \cdots, x_{3m})$, where an element $x_j$ in the sequence is either a trip source ($s_i$), a trip destination ($e_i$), or trip request point ($r_i$). Furthermore, a server is assumed to travel along the shortest path in the road network when moving between any two consecutive points in the trip schedule $x_i$ and $x_{i+1}$.
Thus, the {\em trip cost} between any two points ($x_i,x_j$) in the trip schedule $d_T(x_i,x_j)$ is denoted as 
\[d_T(x_i,x_j)=d(x_i,x_{i+1})+d(x_{i+1,i+2})+\cdots+d(x_{j-1},x_{j}).\]
The overall trip cost is simply $d_T(x_1,x_{3m})$. 

\begin{figure}[bt]
\centering
\includegraphics[width =0.5 \textwidth]{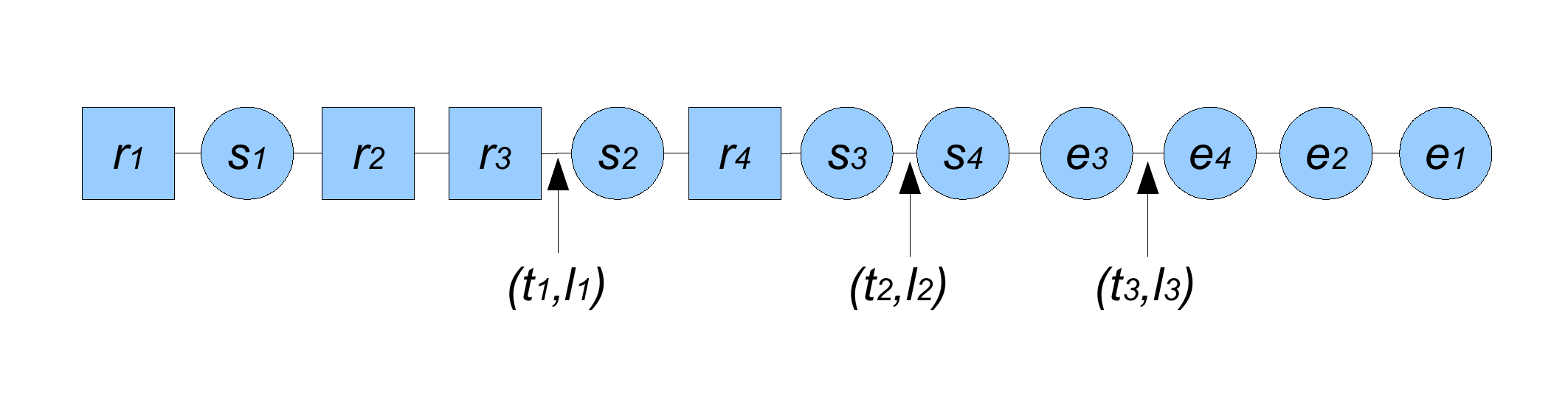}
\vspace*{-5.0ex}
\caption{Trip Schedule. $s_i$: trip starting point; $e_i$: trip ending point; $r_i$ server location when request of trip $tr_i$ comes in; $(t,l)$ is the current time and location of the server.}
\vspace{-0.5cm}
\label{fig:tripschedule}
\end{figure}

Figure~\ref{fig:tripschedule} illustrates a trip schedule for four trip requests.
Note that since each moving server is associated with a trip schedule at any give time, we associate two variables $(t,l)$ with a trip schedule to facilitate our discussion, where $t$ is the current time and $l$ is the current location of the server. 
Intuitively, if a trip schedule is being executed by a server, $(t,l)$ will move along the trip schedule. 
Note that, at any given time $t$, each server is associated with a subset of {\em active trips}, the trips whose requests have been accepted (with some picked up and some not) but not yet dropped off. For instance, in Figure~\ref{fig:tripschedule}, the active trips are $\{tr_1,tr_2,tr_3\}$ at time $t_1$; $\{tr_1,tr_2,tr_3,tr_4\}$ at time $t_2$; and $\{tr_1,tr_2,tr_4\}$ at time $t_3$. 

However, not all the trip schedules can meet the service quality guarantees for each of the individual trip request. We formally introduce the concept of a {\em valid trip schedule}.

\bdefin({\bf Valid Trip Schedule})
A valid trip schedule $S$ for a trip set $TR=\{tr_1,tr_2, \ldots, tr_m\}$ satisfies three conditions: 
\benum
\item{\bf Point order} For any trip $tr_i$, let $x_{i1}=r_i$, $x_{i2}=s_i$, and $x_{i3}=e_i$, then, $i_1 < i_2<i_3$, i.e., the requesting point must happen before the pickup point, which must happen before its ending point; 
\item{\bf Waiting time constraint} For any trip $tr_i$, the distance (waiting time) from the server's location when the request is made to the request's pickup point should be smaller than the waiting time constraint, i.e., $d_T(r_i,s_i) \leq w$; 
\item{\bf Service constraint} For any trip $tr_i$, the actual travel distance from the pickup point to the dropoff point $d_T(s_i,e_i)$ should be smaller than or equal to the shortest distance between them multiplied by the service constraint, i.e., $d_T(s_i,e_i) \leq (1+\epsilon) d(s_i,e_i)$. 
\eenum 
\edefin

%Given a set of vehicles moving on a road network and requests arriving in real-time, for any incoming trip request $tr$, we would like to  quickly identify the vehicles that can (1) combine the request with its existing trips; (2) provide service guarantees for both existing trips and the new request; (3) minimize the total travel time of the vehicle to completely serve both the existing trips and the new request.
%todo: above last part maybe still a bit confusing

To formally define the {\em real-time ridesharing} problem, we
further introduce the {\em augmented valid trip schedule}: {\em Assuming at time $t$, there are $m$ active trips for the given server, let the current valid trip schedule be $(x_1,x_2,\cdots,x_{3m})$, where $(t,l)$ is between $x_i$ and $x_{i+1}$. For a new trip request $tr_{m+1}$, the augmented valid trip schedule shall be $(x^\prime_1,x^\prime_2,\cdots,x^\prime_{3m+3})$, where $x^\prime_j=x_j$ for $j \leq i$, and $x^\prime_{i+1}=r_{m+1}$. } 
In other words, the augmented valid trip schedule combines the new request with the existing requests and shares the same partial trip schedule before the new request is made at time point $t$. Also any augmented valid trip schedule consists of two part: the {\em finished schedule} $(x_1,x_2,\cdots,x_i,r_{m+1})$ and the new {\em unfinished schedule} 
$(r_{m+1}, x^\prime_{i+2},\cdots,x^\prime_{3m+3})$.   

{\em The problem of {\bf Real-Time Ridesharing} is: Given a set of vehicles on the road network $G$ and a new incoming request $tr$, find the vehicle that minimizes the overall trip cost for the augmented valid trip schedule. }

Note that since the finished schedule part in the augmented valid trip schedule cannot be changed (because it has already been executed), we essentially need to find the minimal trip cost for the new unfinished schedule part which includes $m$ active trips and a new trip request. 
We also observe that the minimal cost is for helping determine the best match between the incoming trip request and the available vehicles in a real-time fashion. 
The minimal cost, then, is greedy in nature: When additional new requests comes in, the past optimal choice matching between the current trip request and the server may not be the minimal anymore. 
However, in the real-time request, this type of optimality tends to be the best we can achieve as the future requests are not available and can be rather easily understood and accepted by riders.

Finally, we note that the problem of real-time ridesharing is NP-hard as the classical Hamiltonian path problem can be reduced to this problem (assuming all the trips have the same ending points and requested in almost the same time). For simplicity, the details of NP-hardness is omitted here. 

\comment{
\begin{mydef}[Route of a Trip Sequence]
For a given trip schedule $S= <s_1, \ldots, e_1, \ldots, s_m, \ldots, e_m>$ and a given starting location $s_0$, the shortest path route is defined as $r = <s_0,\ldots, s_1, \ldots, e_1, \ldots, e_m>$ where any consecutive pairs of trip nodes are connected by their shortest path in graph $G$.
\end{mydef}

\begin{mydef}[Route of a Vehicle]
The route of a vehicle $r$ at time $t$ can be represented by $r^t = <0,1, \ldots, m_r^t >$ where $0$ is the node of $G$ the vehicle just visited and $j \in \{1,2, \ldots ,m_r^t\}$ are the nodes of $G$ to be visited in that order. If a vehicle $r$ does not have a predefined route at time $t$, then $r^t = <0,1>$. 
\end{mydef}

Let $R = \{r_1, r_2, \ldots, r_n\}$ be $n$ vehicles.
}

\comment{
\begin{mydef}
A user requests a trip $u^t = <s,e>$ at time $t$ where $u^t.s$ is the starting node and $u^t.e$ is the ending node of the trip and let $d(u^t) = dist_{G}(u^t.s,u^t.e)$ be the shortest distance in terms of time of the trip $u^t$ in graph $G$. At $t$, if there is a valid modification $r_{new}^t$ of a route $r^t$ so that $r_{new}^t$ picks up first and then drops $u^t$ before time  $t + d(u^t) + W$ where $W$ is the system service guarantee, then $r_t$ can absorb $u^t$.  Let $i$ and $j$ be the positions of $u^t.s$ and $u^t.e$ in $r_{new}^t$, then $r_{new}.i$ and $r_{new}.j$ are marked as trip nodes and a waste time $w=dist_{r_{new}^t}(r_{new}^t.0,r_{new}^t.j) - d(u^t)$ is associated with $r_{new}^t.i$ and $r_{new}^t.j$ (denoted by $r_{new}^t.i.w$ and $r_{new}^t.j.w$). It is obvious that $r_{new}^t.i$ and $r_{new}^t.j$ are positive numbers initially.
\end{mydef}

\begin{mydef}
Let $S(r)$ be a function that extracts the subsequence of trip nodes from a route $r$.
A modification of $r^t$ to $r_{new}^t$ is valid, if (1) $S(r^t)$ is a subsequence of $S(r_{new}^t)$; (2) For any trip $u'$ with $u'.e$ at position $i$ of $r^t$ and $j$ of $r_{new}^t$, $d = dist_{r_{new}^t}(r_{new}^t.0, r_{new}^t.j) - dist_{r^t}(r^t.0, r^t.i) \le r^t.i.w$ and $r_{new}^t.j.w$ is set as $r^t.i.w - d$. Let $l$ be the location of $u'.s$ in $r_{new}^t$, then $r_{new}^t.l.w$ is set as $r^t.i.w - d$ as well.
\end{mydef}

Let $S(r)$ to be a function that extracts the subsequence of trip nodes from a route $r$. For all the routes at $t$ that can absorb a user request $u^t$.

\begin{mydef}
For a route $r^t$ and a user request $u^t$, all the valid modifications of $r_t$ result in multiple routes $R_r^t=\{r_1^t, \ldots, r_k^t\}$.  Then the sequences in $R_r^t$ can form a tree structure $T_r^t$ called route tree by combining common prefix. When the vehicle reaches one node in its route at time $t$,  its current route tree is updated by deleting the first node in the route tree together with all its children whose root is not the next node that the vehicle is going to.
\end{mydef}

Let $S(r)$ be a function that extracts the subsequence of trip nodes from a route $r$. Let ${\cal R}_r^t=\{S(r)| r \in R\}$.

}

\subsection{Challenges}
The main challenge in ridesharing is to determine how to handle trip requests as they flow into the system in real-time. From a server's point of view, for any new request, each server may have already selected (and be executing) a trip schedule for its existing customers. 
Given this, how can we quickly help it to determine whether it can accommodate a new request?  Note that in order to respond to such a request, one may have to reshuffle the predefined schedule and the reshuffled one has to be a valid schedule.

Furthermore, there might be tens to even hundreds of servers in the region surrounding the pickup point of a new request. Clearly, for a trip request $tr_i$, servers that are farther than $w$ from the pickup location are unable to respond to the request. Thus, we can already reduce the potential candidates to only those that are within $w$ of the pickup point. Then, the customer will be assigned to the server that offers the shortest total trip time. Even though potential servers can be filtered through a dynamic spatial indexing structure \cite{Tao:2003, Mokbel:2004, Jensen:2004} on the moving servers, the existing approaches can still be very computationally expensive and result in low response times. In a large metropolitan area such as Shanghai, the number of requests can be very large, especially during rush hour. 

Most algorithms are designed for offline computation. The existing approaches that use branch-and-bound \cite{Kalantari:1985} or integer programing \cite{Cordeau:2006} to schedule new requests do not take the dynamic nature of the problem into consideration. Testing if a new request can be accommodated essentially involves a rescheduling of the unfinished trips and the new request without reusing the computations in the previous round. Their calculation time was measured in minutes or hours while we require milli-second response time.
% minutes or hours???

\subsection{Contributions}
To deal with the challenges, our idea is based on a simple observation. 
For a new valid schedule accommodating the new request $tr_i$, if we simply drop the three points $r_i, s_i, e_i$ from the trip schedule, then the resulting trip schedule is a valid trip schedule. In other words, only a valid trip schedule can be extended to accommodate a new request. Given this, a potential approach for the ridesharing problem is to simply materialize every valid trip schedule; then, when a new request arrives, we can check if any valid trip schedules can be extended to handle the new request. This approach is promising because its incremental nature saves many redundant computations: We do not need to recompute the valid trip schedule completely from scratch on each new request. 
However, in order to implement such a strategy, we have to deal with the following challenges: 
1) Would the materialization incur too much memory cost? In other words, can we store the materialized schedules compactly?
2) How can we efficiently maintain the materialization? Note that when the server moves, the materialization needs to be updated. 
3) How can the materialization help to test quickly whether a new request can be handled?
4) How can the materalizaton be updated when a new requested is accepted?

%To deal with these challenges, we introduce a new data structure, referred to as a {\em Kinetic-Valid-Schedule-Tree}, to handle the materialization and to help quickly respond to requests. 

This paper makes the following contributions:
\begin{itemize}
\item We formulate the ridesharing problem in a way that resembles the scenario enabled by current locating and communication technology; We propose a kinetic tree approach for the matching problem. The tree structure lends itself naturally to the dynamic nature of the problem;
\item When the pickup or dropoff locations are close to each other, any permutation of the locations can be valid, rendering the constraints ineffective and resulting in a large number of valid schedules. We propose a hotspot-based algorithm that ignores schedules that are almost duplicates to effectively reduce the number of valid schedules while providing a bound on the error for the solution under certain conditions;
\item We compare our approach to the branch-and-bound and mixed integer programing approaches that are traditionally used together with the brute-force algorithm. Experiments on a large shanghai dataset show that the tree algorithm is several times to a magnitude faster in response time. We further test tree algorithm on various larger problems to show its performance and effectiveness of the optimizations proposed.
\end{itemize}

\subsection{Outline}
We first describe a branch-and-bound an mixed-integer programming algorithm to solve the problem in Section \ref{tradition}. We then propose the kinetic tree approach in Section \ref{tree}. In section \ref{large}, we deal with the issue of large trees using a hotspot-based algorithm. Experiment results are presented in Section \ref{experiment}. we discuss related work in Section \ref{relate}, and present our conclusions in Section \ref{conclusion}.

\section{Branch-and-Bound and Mixed Integer Programming Algorithms} \label{tradition}
The brute-force algorithm to find the augmented valid trip schedules is straightforward. We enumerate all of the permutations and then check the constraints. However, this can be expensive. Two traditional approaches that are often used in solving the related dial-a-ride problem \cite{Feuerstein:2001} can increase execution speed: a branch-and-bound algorithm \cite{Colorni:2001} and an integer programming approach \cite{Cordeau:2006}. We first propose a modified branch-and-bound algorithm for our problem, and then formulate the problem as a mixed-integer programming problem. 

\section{Branch and Bound Algorithm}

The branch-and-bound algorithm systematically enumerates all candidate schedules and organizes the candidates into a schedule tree. It estimates and maintains a lower bound of each partially constructed schedule and stops building candidate schedules that have lower bounds greater than the best solution found so far. The algorithm first expands the partial candidate with the lowest lower bound (best first search).

\begin{figure}[bt]
\centering
\includegraphics[width =0.5 \textwidth]{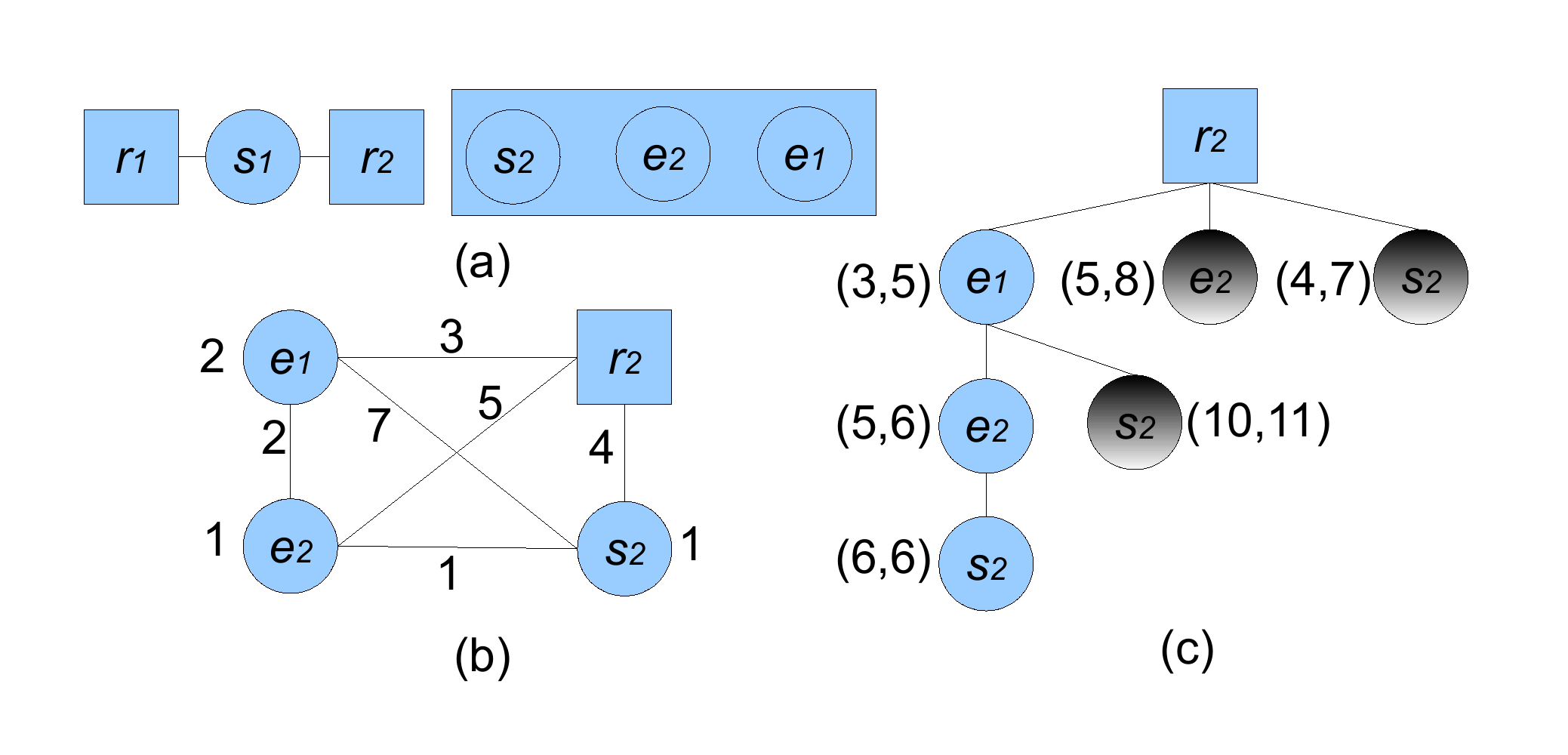}
\vspace*{-5.0ex}
\caption{Illustration of Branch-and-Bound Algorithm. (1) When request $r_2$ comes, only $\{e_1,s_2,e_2\}$ need to be scheduled; (b) Road network distance and minimal incident edge cost; (c) When $(r_2,e_1,e_2,s_2)$ with cost 6 is found, partial schedules with estimated costs above 6 are terminated.} 
\vspace*{-0.3cm}
\label{fig:BB}
\end{figure}

Assume at time $t$, there are $m$ active trips for the given server. Let the current valid trip schedule be $(x_1,x_2,\cdots,x_{3m})$, where $(t,l)$ is between $x_i$ and $x_{i+1}$. For a new trip request $tr_{m+1}$, we need to re-schedule the  pickup and dropoff points $N=\{x_{i+1}, x_{i+2}, \cdots, x_{3m},$  $ r_{m+1},s_{m+1}, e_{m+1}\}$.  We treat $N$ as a complete graph with vertices being $N$ and edge weights being the shortest path distances between nodes. We attempt to find the schedule through the graph that passes through each node once but, unlike a tour, does not return to the first node. The schedule also has to begin at the location of the server $l$ (this is also $r_{m+1}$). In Figure \ref{fig:BB} (a), when request $r_2$ comes in, $s_1$ is already picked up. So, only $N=\{e_1,s_2, e_2\}$ needs to be scheduled and the schedule must start from $r_2$.

%Algorithm \ref{alg:branch} applies the branch-bound method on the schedules through the desired pickup and dropoff locations of a trip set $TR=\{tr_1,tr_2,\ldots, tr_m\}$. We treat the set of starting nodes and ending nodes $TN(TR)$ as a complete graph and attempt to find the schedule through the graph that passes each node once but, unlike a tour, does not return to the first node in the sequence. The schedule also has to begin at the original location of the server $r$.

%However, because this algorithm is executed each time a trip request is received, some of the pickup and dropoff locations may have already been passed. Then, only a set of $n$ locations $N = \{N_1, ..., N_{n}\} \subseteq TR$ need to be scheduled with the current location $r$ of the serving vehicle when the trip request is received. Although the vehicle already has a best schedule, the request adds two new nodes that may change the schedule. 

We start with the initial schedule tree $ST = <r_{m+1}>$, and initialize the cost of the optimal schedule to $\infty$. We then iteratively perform a best-first-search to expand the partial schedule $S = <r_{m+1},x^\prime_{i+1}, x^\prime_{i+2}, \cdots, x^\prime_{k}>$ with the minimum lower bound. The key to a branch-and-bound algorithm is to find an effective lower bound. The bound we use is $d_T(r_{m+1},x^\prime_{k})$ plus the sum of the minimum-cost-edge incident to the nodes that are not yet in the partial schedule $S$. 

Figure \ref{fig:BB} (b) shows road network costs between two nodes. The minimal incident edge cost is labeled beside each node. In Figure \ref{fig:BB} (c), for each node $x$, the two numbers in a parentheses indicate the cost $d_T(r_2,x)$ of the partial schedule and the lower bound of the schedule containing the partial schedule as prefix. For $(r_2,e_1)$, $d_T(r_2,e_1) = 3$. Only $e_2$ and $s_2$, both with minimal incident edge cost of 1, need to be added to the schedule, so the lower bound of a schedule containing $(r_2,e_2)$ is $d_T(r_2,e_2) + 1 + 1 = 5$.   %Algorithm \ref{alg:bound} shows how the boundes are calculated.

%\begin{algorithm}
%\fontsize{9}{9}\selectfont
%\begin{algorithmic}
%\REQUIRE a path $p$ and set of nodes $N$
%\STATE Initialize $bnd = p.length$
%\FORALL{$n \in N - p$}
%	\STATE $bnd = bnd + \text{minedge}(n)$ \COMMENT{minedge returns the weight of the minimum edge connected to $n$}
%\ENDFOR
%\end{algorithmic}
%\caption{Bounding algorithm for a path $p$.}
%\label{alg:bound}
%\end{algorithm}

We attempt to expand the partial schedule $S$ with minimal lower bound by another new node to construct $S'$. If $S'$ is not valid or results in a bound greater than the current minimum schedule cost, we terminate $S'$. If $S'$ is a complete schedule, we compare its cost to that of the best schedule and update if necessary. Figure \ref{fig:BB} (c) shows the execution figure \ref{fig:BB} (a) . Once the schedule of cost $6$ is found, schedules with lower bounds above 6 can be pruned (labeled by a gray circle). Note that in the figure we do not illustrate validity constraints. The complexity of the branch-and-bound algorithm in the worst case is still exponential. 

\comment{
\begin{algorithm}
\fontsize{9}{9}\selectfont
\begin{algorithmic}
\REQUIRE current location $r$ and set of nodes $N$
\STATE Initialize minimum tour cost $mt = \infty$
\STATE Initialize initial path $p_{initial} = <r>$
\STATE Initialize $p.cost = 0$ \COMMENT{Store $d_T$ as $p.cost$ so it doesn't need to be recalculated}
\STATE Initialize priority queue $Q = \emptyset$ of paths, where the priority of an element $x$ is $\text{bound}(x)$
\STATE Insert $p_{initial}$ into priority queue $Q$
\WHILE{$Q$ is not empty}
	\STATE $t = Q.\text{pop}()$
	\IF{$\text{bound}(t) < mt$}
		\FORALL{$x \in N - t$}
			\STATE $u = t \cup x$
			\STATE $u.cost = t.cost + \text{d}(u_{|u| - 1}, u_{|u|})$
			\IF{$u$ is valid}
				\IF{$|u| = |N| - 1$}
					\STATE Append the only node $y \notin u$ to $u$
					\STATE $u.cost = u.cost + \text{d}(u_{|u| - 1}, u_{|u|})$
					\IF{$u.cost < mt$}
						\STATE $mt = u.cost$
					\ENDIF
				\ELSIF{$\text{bound}(u) < mt$}
					\STATE $Q.\text{push}(u)$
				\ENDIF
			\ENDIF
		\ENDFOR
	\ENDIF
\ENDWHILE
\end{algorithmic}
\caption{Branch and bound algorithm.}
\label{alg:branch}
\end{algorithm}
}

\subsection{Mixed-integer Programming Approach}
\label{LP}
Mixed integer programing is a popular alternative. In this section, we formulate our augmented valid trip schedule problem into a mixed integer programming problem. Then, we apply traditional solvers to find the solution. 

As in the branch and bound algorithm, we are rescheduling $N=\{x_{i+1}, x_{i+2}, \cdots, x_{3m}, r_{m+1},s_{m+1}, e_{m+1}\}$. The schedule must start from $r_{m+1}$. We divide $N$ into subsets: (1) dropoff locations of those already picked up but not dropped off; let the size of this set be $k$; (2) pickup locations of trips not started yet; let the size of this set be $n$;  and (3) dropoff locations of trips not started yet; the size of this set is also $n$. The problem can be defined on a complete directed graph $G=(N,A)$ where $N=D' \cup P \cup D \cup \{0\}$, $D'=\{1,2,\ldots, k\}$, $P=\{k+1,k+2,\ldots, k+n\}$, $D=\{k+n+1,k+n+2,\ldots,k+2n\}$. Here we assign an integer to each point in $N$ while node 0 represents the current position $l/r_{i+1}$ of the server. For a pickup $i$ in $P$, its matching dropoff in $D$ is $i+n$. A pickup constrain  $l_i$ is associated with a node $i\in P$, representing the latest time that node $i$  need to be picked up. Each arc $(i,j) \in A$ are associated with a shortest path routing cost $d_{ij}$. For each arc $(i,j)$, let $y_{ij} = 1$ if the server travels from node $i$ to node $j$. For each drop point $i \in D' \cup D$, let $L_i$ be the ride time of the request with dropoff $i \in D' \cup D$ in this partial route.

$$
Min \sum_{i\in N}\sum_{j \in N} {d_{ij}y_{ij}}
$$

subject to:

\begin{center}
\begin{tabular}{llll}
$y_{ij} \in \{0,1\}$, &$\forall i \in N, j \in N$ & (1)\\
$\sum_{j\in N} {y_{ji}} = 1$, &$\forall i \in N - \{0\}$ & (2)\\
$\sum_{j\in N} {y_{0j}} = 1$ && (3)\\
%$\sum_{i\in N}\sum_{j \in N}{x_{ij}} = m+2n$ && (4)\\
$B_0 = 0$ && (4)\\
$B_j \geq (B_i + d_{ij}) y_{i,j}$, & $\forall i \in N, j\in N$ & (5) \\
%$L_i = B_{i}$, & $\forall i \in D'$ & (6)\\
$L_i = B_{i} - B_{i-n}$, & $\forall i \in D$ & (6)\\
$B_i \leq w_i$, & $\forall i \in P$ & (7) \\
$B_i \leq r_i$, & $\forall i \in D'$ & (8) \\
$d_{i-n,i} \leq L_i \leq \epsilon_i$, & $\forall i \in D$ & (9) \\
%$t_{i,n+i} \leq L_{n+i}$, & $\forall i \in P$ & (5)\\
%$t_{0,i} \leq L_{i}$, & $\forall i \in D'$ & (6)\\
\end{tabular}
\end{center}

where $w_i$ is the waiting time left for $i \in P$ and $r_i$ is the maximal riding time left for $i \in D'$. Here $d_{ii}$ is set to a positive number to make sure $y_{ii} = 0$.

The objective is to find the schedule that minimizes the total cost while satisfying the constraints. Constraint (1) simply enforces the binary nature of $y_{ij}$. Constraint (2) allows exactly one node preceding another for all nodes but $0$. Constraint (3) allows exact one node following node $0$. These two effectively enforce the schedule structure so that each node is visited exactly once and the schedule starts from node 0. %In other words, other than the first and the last node, every node has exactly one node preceding it and one following it. We cannot require a node to have one node to follow it because the last request node ends completes the path and this node is variable.

Constraints (4) and (5) set the earliest time at which a node can be reached. Constraints (6) define $L_i$ for dropoff nodes, the service distance. Constraints (7) and (8) enforce the waiting time and service constraints for pickup and dropoff nodes where the passenger has already been picked up. These are grouped together because both $w_i$ and $r_i$ are measured from the root node. Constraint (9) enforces the service constraint for dropoff nodes where the passenger has not yet been picked up, so that the service time does not exceed $\epsilon_i$.

The constraint (5) is not linear. It can be linearized by introducing constants $M_{ij}$ using the idea similar to the Miller-Tucker-Zemlin subtour elimination constraints for the traveling salesman problem \cite{Desrochers:1991}:
\begin{equation}
\centering B_j \ge B_i + d_{ij} - M_{ij}(1-y_{ij}), \forall i \in N, j \in N
\end{equation}

The validity of these constraints are ensured by setting $M_{ij} \ge \max\{0,l_i+d_{ij}-e_j\}$ where $l_i$ is the latest time that $i$ need to be served and $e_j$ is the earliest time that $j$ needs to be served. For $i \in P$, $[e_i,l_i] = [d_{0i},w_i]$. For $i \in D, [e_i,l_i]=[d_{0,i-n} + d_{i_n,i}, w_i + d_{i-n,i} (1+\epsilon)]$. For $i \in D', [e_i,l_i] = [d_{0i},r_i]$.

%\subsubsection{Complexity Analysis}

Let $v$ be the number of variables in the mixed-integer programming problem, and $c$ be the number of constraints, then $v = O(m^2)$ and $c = O(m)$, where $m$ is the total number of requests that we are optimizing. %Based on this, it can be seen that the mixed-integer approach is theoretically infeasible with large vehicle capacities (which serve as a bound on $m$) as a result of the $m^2$ relation. Nevertheless, this may not be a problem because practically there is a limit on the vehicle capacity. 

%Thus, in this experiment we are more concerned with speed on solving small problems.
%Although buses may be able to accomodate as many as 300 passengers, most vehicles will have a much fewer passengers and public transportation systems rarely are able to hold more than 300 in a single vehicle. Thus, in this experiment we are more concerned with speed on solving small problems.

\section{Kinetic Tree Approach}
\label{tree}
The two approaches above both suffer from one fundamental problem: they essentially reschedule the unfinished pickups and dropoffs with the new request from scratch without re-using the computations performed before. The structure of the two algorithms make it difficult to adapt to the dynamic nature of the problem. In this section, we introduce a kinetic tree structure that can maintain and update the calculation performed up-to-now and use them effectively when a new request is issued. However, when there are multiple pickup or dropoff locations close to each other, the possible number of schedules inevitably increase in an exponential fashion. We then propose a hotspot-based approach in section \ref{large} that reduces the search space and approximates the solution with bounds.

%To be fair and speedup the response time, we may want to all server driver who can actually accommodate such a new request to get notified quickly and around the same time. 

\begin{figure}[t]
\centering
\includegraphics[width =0.5 \textwidth]{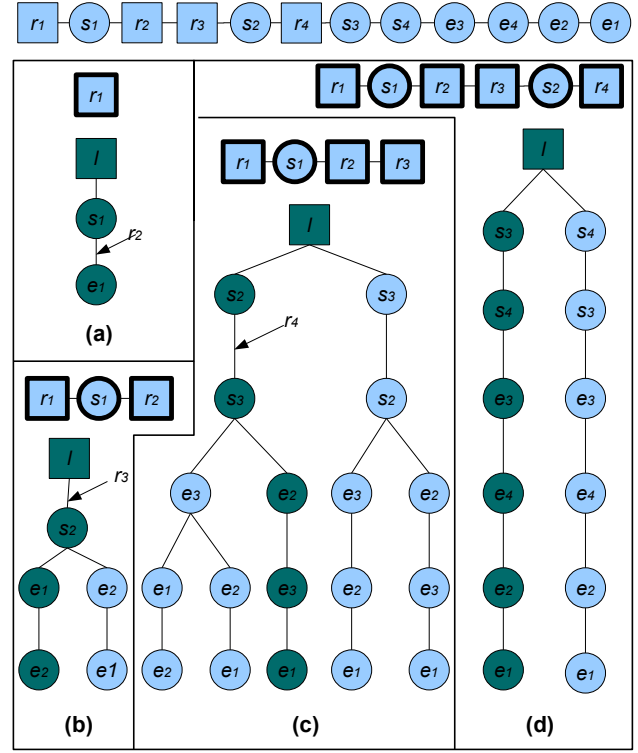}
\vspace*{-10.0ex}
\caption{Kinetic Tree for Trip Schedules. Darkened path: selected schedule to be executed; Dark circled/squared nodes: finished nodes.}
\label{fig:tripscheduletree}
\vspace*{-0.5cm}
\end{figure}

\subsection{Basic Tree Structure}
\label{basic}
We introduce a kinetic tree structure to maintain all the valid trip schedules with respect to the server's current location. When the server moves, a portion of the schedule becomes obsolete. The root of the tree tracks the current location $l$ of the server. The rest of the tree records the portion of all the valid schedules (from the current location onwards).

For a given $w$ and $\epsilon$, Figure~\ref{fig:tripscheduletree} illustrates the kinetic tree structure corresponding to the complete trip schedule in Figure~\ref{fig:tripschedule}. The darkened path represents the selected schedule to be executed by the server.  Initially, for the first trip request, there is only one valid trip schedule (Figure~\ref{fig:tripscheduletree} (a)). 
When the second request arrives, the first customer has already been picked up by the server. 
Now, consider there are only two valid options for the server to accept the new request: it needs to first pick up the second customer, but it can be flexible in dropping off either of the two passengers. Let us assume it decides to choose the shorter one which is $(l,s_2,e_1,e_2)$, to drop off the first customer first (assuming the option which drops the first passenger before picking up the second one is invalid). However, on its way to pick up the second customer, the third request arrives. The server now has the options to either pick up the second customer or the third one.  Suppose, consequently, based on $w$ and $\epsilon$, that there are five possible valid trip schedule for the server to handle the three trip requests (trip one is already in progress, shown in 	Figure~\ref{fig:tripschedule}(c)). Assuming the server decides to move along the shortest route $(l,s_2,s_3,e_3,e_3,e_1)$ for now and picks up the second customer first, then when the fourth request arrives after the pickup of the second customer, the entire right subtree of $r_3$ in Figure~\ref{fig:tripschedule}(c) becomes inactive. Let us now assume there are only two possible schedules to accommodate the remaining trips of all four customers as shown in  Figure~\ref{fig:tripschedule}(d).

Why is such a kinetic tree useful in maintaining the valid trip schedules? 
Its advantage is based on the the following key observation: 

\begin{lemma}[Valid Schedules under Movement]
When a server reaches a new pickup location or dropoff location in the trip schedule, then only those valid schedules which contain unfinished trips and share the same prefix so far (from the first pickup point of all the unfinished schedules to the current location in the trip schedule) need to be materialized. All the other schedules will become inactive and can be pruned from the tree. 
\end{lemma}

For example, in Figure~\ref{fig:tripscheduletree}(c), once the server actually picked up the second customer, only the schedules in the left subtree rooted with $s_2$ remain active. 
%Note that after server leaves a pickup/dropoff location, on its way, some branches (trip schedules) may become inactive before it actually reaches the next location (the root node of its targeted subtree). For instance, in Figure~\ref{fig:tripscheduletree}(c), before the server reaches $s_2$ (the root node of its left subtree), all the trip schedules in the right subtree have already become inactive once we know the server is not on the shortest path towards $s_3$. 
There are two options to perform the tree pruning. 
The {\em eager invalidation} option tries to determine whether some trip schedules become inactive as early as possible. 
In other words, it tries to perform the pruning as the server moves or when the next point in the scheduled route is reached. The {\em lazy invalidation} option only performs such pruning when necessary, i.e., only when there is a new incoming request. 
%In the latter case, all the rest of the subtrees can be pruned. 

\subsection{Handling a New Request}
\label{newrequest}
Now, we consider how to handle a new request $tr_k=(r_k, s_k,e_k)$. 
The assumption is that we already have a materialized prefix tree of all valid and active  schedules of unfinished trips. Now, we need to extend all valid and active schedules in the prefix tree to a new valid schedule to include $tr_k$ if possible. We do this by generating a new prefix tree based on the existing one. To deal with the new request, we will first deal with the pickup location $s_k$ and then the dropoff location $e_k$. Essentially, we need to scan the tree to determine where $s_k$ can be inserted, i.e., which edges of the tree can accommodate the insertion of a new pickup node. All schedules that share the prefix from the root of the tree to the inserted edge will be inserted into the new tree. Then we insert $e_k$ after $s_k$ in the new tree. Furthermore, if $s_k$ or $e_k$ can be inserted at a given location (an edge in the tree), then we have to find out which trip schedules containing that edge with an additional node will become invalid and should be pruned from the new tree. The problem is how to determine 1) at which edge $s_k$ or $e_k$ can be inserted, and 2) how to quickly prune the invalid trip schedules following that insertion.\\ 

\noindent{\bf Inserting Pickup Location:}
Here, we focus on whether $s_k$ can be inserted first and $e_k$ can be inserted in a similar way later. In order to insert $s_k$ in a tree edge, say $(x_i,x_{i+1})$, we need to deal with the following situations: 
(a) only when the distance from the current location (recorded in the root node $l$) to the pickup location $s_i$ satisfies  $d_T(l,s_i)=d(l,x_1)+d(x_1,x_2)+\cdots+d(x_i,s_k) \leq w$, then $s_k$ may be inserted; 
(b) the additional travel distance (time) introduced by the detour to $s_k$ may invalidate some existing trip schedule in the subtree containing this tree edge $(x_i,x_{i+1})$, i.e., $d(x_i,s_k)+d(s_k,x_{i+1})-d(x_i,x_{i+1})$ should not be too large. These schedules should be pruned from the subtree. Note that condition (a) is easy to be tested in the existing tree structure. 

\blemma({\bf $d_T(l,s_k) \leq w$}) 
The shortest distance from the current location to the requested pickup location $s_k$ is no larger than $w$. 
Furthermore, given a prefix (partial) trip schedule from the root node $l$ to a node $x_j$, i.e., $(l,x_1,x_2, \cdots, x_j)$, 
if $d_T(l,s_k)=d(l,x_1)+d(x_1,x_2)+\cdots+d(x_j,s_k) > w$, then, any edge incident to any descendant of $x_j$ in the tree cannot accommodate $s_k$, i.e., customer can not wait for server until it finishes $x_j$ to pick him up at $s_k$.   
\elemma

This lemma suggests that we can perform either a depth first search (DFS) or breadth first search (BFS) starting from the root node of the tree to generate all the candidate edge $(x_i,x_{i+1})$ to insert $s_k$. Specifically, during the traversal the visiting will return once certain depth is reached, i.e., a node has the property that $d_T(l,x_j) >w$, then, we either will not expand that nodes (in BFS) or trace back (in DFS).

Now the key problem is how to handle  condition (b).
The straightforward way to perform pruning is to explicitly maintain and check the constraints for each trip request in the subtree of the node $x_i$. Specifically, for a trip $tr_j$ in the subtree rooted at $x_i$, there are two criteria: pickup waiting constraint $[r_j,s_j,w]$ ($d_T(r_j,s_j)\leq w$) and trip tolerance constraint $[s_j,e_j,\epsilon]$ ($d_T(s_j,e_j) \leq (1+\epsilon) d(s_j,e_j)$). 
At any given time point $t$, clearly if we need to test whether the detour meets the criteria of trip $tr_j$, then the request is already issued and responded, and the entire trip is not yet completed. Further more, only one of the criterion needs to be tested: if the server has not picked up the customer, then, we need to test the pickup waiting constraint $[r_j,s_j,w]$; once the customer is picked up, we need test the trip tolerance constraint $[s_j,e_j,\epsilon]$. 
Thus, at any given point, the ``active'' customers can be partitioned into two sets: $S_1$ records those customers who need to be picked up and $S_2$ records the on-board customers who need to be dropped off. When a new location is reached, we may move customers from $S_1$ to $S_2$ and/or remove customers from $S_2$. 
For trip $j$ in $S_1$, we test the first criterion $[r_j,s_j,w]$ and in $S_2$, we test the second one: $[s_j,e_j,\epsilon]$. 
Given this, for the subtree rooted at $x_i$, the straightforward way is to first generate these two sets $S_1$ and $S_2$. Then, when we insert $s_k$, we need test each condition associated with $S_1$ and $S_2$ are also satisfied. \\

\begin{algorithm}
\fontsize{9}{9}\selectfont
\begin{algorithmic}
\REQUIRE root node $l$, request points $P = (x_1, x_2, ...)$, current depth $depth$
\IF{$feasible(l, x_1, depth + d(l, x_1))$}
	\STATE Initialize $fail = 0$
	\STATE $n = create(l, x_1)$
	\COMMENT{Copy feasible child branches underneath $n$}
	\FORALL{$c$ such that edge $(l, c)$ exists}
		\STATE $copyNodes(n, \{c\}, d(l, n) + d(n, c) - d(l, c))$
		\STATE If copy failed, set $fail = 1$
	\ENDFOR
	\COMMENT{Insert remaining request points to $n$}
	\IF[Detour now begins negative because we haven't inserted $x_2$ yet]{$fail = 0$ and $|P| > 1$}
		\STATE $insertNodes(n, \{x_2, ...\}, -d(x_1, x_2))$
		\STATE If insert failed, set $fail = 1$
	\ENDIF
	\COMMENT{Now insert request points into children}
	\FORALL{$c$ such that edge $(l, c)$ exists}
		\STATE $insertNodes(c, P, detour + d(l, c))$
		\STATE If insert failed, delete $(l, c)$
	\ENDFOR
	\IF{$fail = 0$}
		\STATE Add edge $(l, n)$
	\ELSIF{No nodes $c$ with edge $(l, c)$ exist}
		\STATE Insert failed, notify caller that this subtree is infeasible
	\ELSE
		\STATE Insert succeeded
	\ENDIF
\ELSE
	\STATE Insert failed, notify caller that this subtree is infeasible
\ENDIF
\end{algorithmic}
\caption{insertNodes algorithm.}
\label{alg:insert}
\end{algorithm}

Algorithm \ref{alg:insert} implements the insertion of a new request $tr_k = (s_k, e_k)$ into the tree recursively. The insertion is completed by a call, $insertNodes(root, \{s_k, e_k\}, 0)$. The call to $feasible(parent, node,$ $ detour)$ returns whether or not it is feasible to insert $node$ as a child under $parent$ in the tree. First, this ensures that the pickup or service constraint of $node$ is not violated. If min-max filtering is in place (will be discussed in next section), this will confirm that the detour (third argument) is tolerable for $node$.

\begin{figure}[bt]
\centering
\includegraphics[width =0.5 \textwidth]{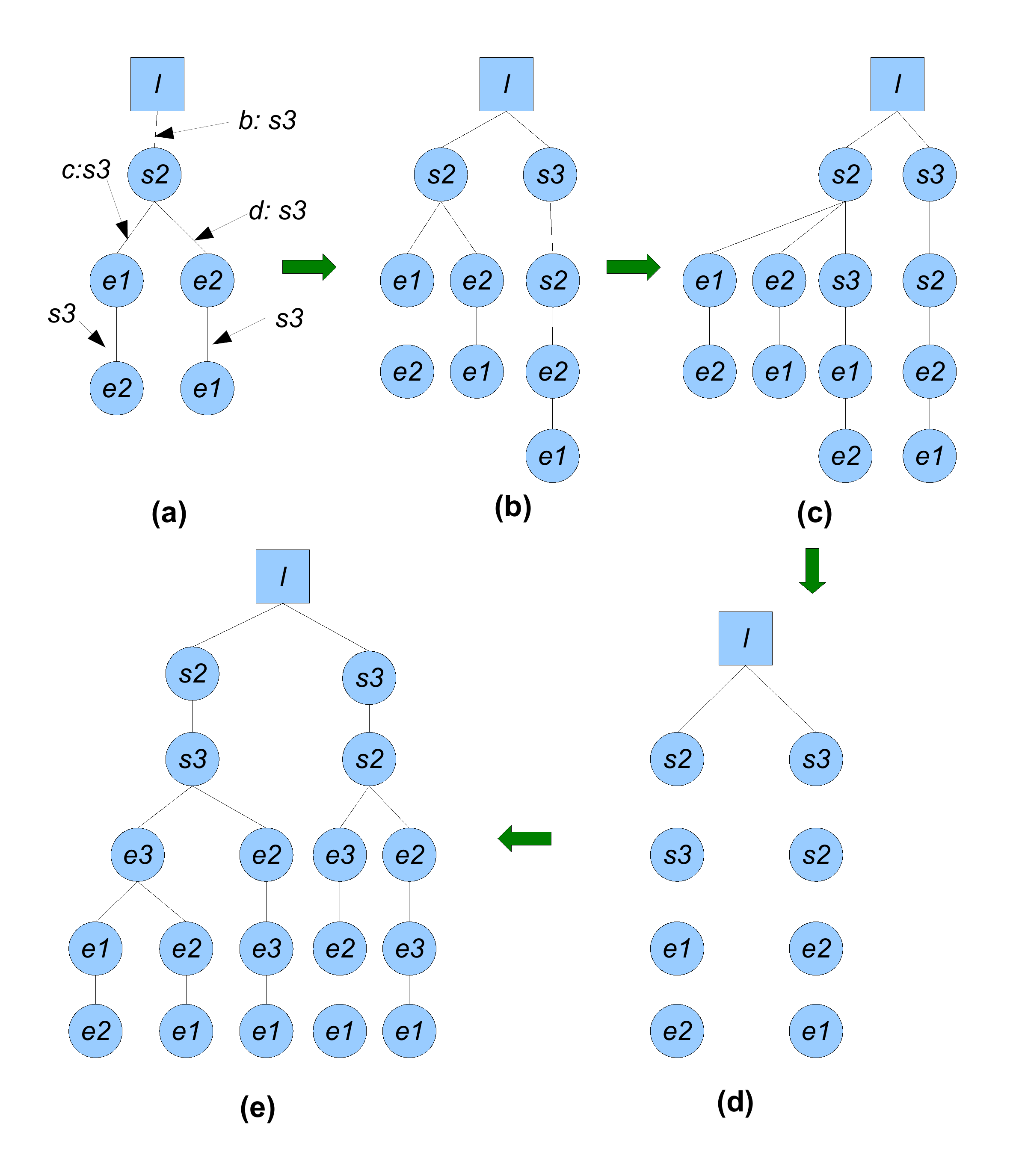}
\vspace*{-7.0ex}
\caption{Tree Insertion. The insertions of $s_3$ into each edge in tree of (a) result in new trees in (b), (c), (d), and (e), assuming the last two insertions were infeasible.}
\label{fig:tripscheduletreeinsert}
\end{figure}

The $copyNodes(node, source, detour)$ function recursively copies nodes from a set of nodes, $source$, to the target node, $node$. Here, tolerance of the root's children in $insertNodes$ is implemented through calls to $feasible$ with detour of $detour$. $copyNodes$ will fail if all of the children of $node$ are along infeasible paths. In this case, these branches and $node$ will be deleted.

In Figure \ref{fig:tripscheduletreeinsert} (a), we use the insertion algorithm to insert the pickup location $s_3$ into an existing tree, thereby generating a new tree. $s_3$ will first be inserted directly below $r$. Then, the branch with root at $s_2$ will be copied underneath this new $s_3$ node, forming a new tree of $(l,s_3,s_2,((e_1,e_2),(e_2,e_1)))$ is generated. Let us assume route $(l,s_3,s_2,e_1,e_2)$ is not feasible; then, the branch is pruned from the tree starting at the leaf node until we reach $s_2$, where we have an alternate path $l,s_3,s_2,e_2,e_1)$ that is feasible. This pruning occurs in the $copyNodes$ algorithm, which will succeed because $s_3$ falls along at least one (in this case, exactly one) feasible path. The resulting tree is shown in Figure \ref{fig:tripscheduletreeinsert} (b).

Then, the insertion algorithm moves down to $s_2$ and attempts to insert the pickup location after it. Two paths are formed: $(l,s_2,s_3,e_1,e_2)$ and $(r,s_2,s_3,e_2,e_1)$, as a result of the insertion between $s_2$ and $e_2$ and between $s_2$ and $e_1$. The resulting tree is shown in Figure \ref{fig:tripscheduletreeinsert} (c). Suppose inserting $s_3$ between $e_1$ and $e_2$ or between $e_2$ and $e_1$ is infeasible. Then, we have the tree in Figure \ref{fig:tripscheduletreeinsert} (d). To complete the insertion of the $(s_3, e_3)$, we now try to insert $e_3$ in the subtrees that root at a $s_3$ following the same insertion algorithm. Once this completes, we arrive at the tree shown in Figure \ref{fig:tripscheduletreeinsert} (e).\\

\noindent{\bf Min-max Filtering using Slack time:}
Though the above test for condition (b) is conceptually simple, it is rather computationally expensive. 
Now, we introduce a fast approach to simplify and speedup such test. 
For any node $j$, if $j$ is in $S_1$, let $\delta_j= w -d_T(r_h,s_j)$; otherwise ($j$ is in $S_2$), let $\delta_j=(1+\epsilon)d(s_j,e_j)-d_T(s_j,e_j)$. 
Then, for the node $x_j$, we associate {\bf slack time} $\Delta_{x_j}=\min (\delta_j,\max_{i \in x_j.children} \Delta_i)$.

Note that $\Delta_{x_j}$  essentially represents the minimal allowed detour on the most ``lenient'' route of the subtree routed at $x_j$. Here ``lenient'' means the route can tolerate the most detour compared to other routes. Given this, we introduce the following Theorem to describe the simple condition to determine whether $s_k$ can be inserted at a given edge. 

\bthm 
For a trip request $tr_k$, if edge $(x_i,x_{i+1})$ does not satisfy either of the following condition: 
(a) $d_T(l,s_k)=d(l,x_1)+d(x_1,x_2)+\cdots+d(x_i,s_k) \leq w$; 
or (b) $d(x_i,s_k)+d(s_k,x_{i+1})-d(x_i,x_{i+1}) \leq \Delta_{(x_i,x_{i+1})}$, 
then, we can not add the pickup $s_k$ between location $x_i$ and $x_{i+1}$.\\
\ethm

%\noindent{\bf Pruning:}
%When not filtered, the insertion will generate a new ``subtree'' that includes a prefix $(l, \ldots, x_i,s_k, x_{i+1})$ and a copy of the original subtrees of $x_{i+1}$. The new subtree is inserted into the new tree being created. 

After insertion in $(x_i,x_{i+1})$, the all nodes under $x_{i}$ of the new tree will be tested for the constraint $\delta_i \ge d(x_i,s_k)+d(s_k,x_{i+1})-d(x_i,x_{i+1})$ . A branch is pruned from the subtree if the constraint is not satisfied. \\ %The $\Delta$ values of all the nodes in the tree rooted at $x_i$ are decreased by the extra detour time $d(x_i,s_k)+d(s_k,x_{i+1})-d(x_i,x_{i+1})$.

%Note that this result simply help determine if we can detour to the pickup location and does not confirm whether the we can detour to the dropoff location. However, the handling of the dropoff after the insertion of the pickup location is quite similar to this one and happens only in the newly created subtree.

%\subsection{Insertion Algorithm}
%\label{algorithm}

\noindent{\bf Updating $\Delta$ and Tree:}
After we try to insert a request to all possible servers, we get a set of new trees. For each tree, we can find the shortest route and choose the tree what provides the shortest route among all trees. Only the chosen tree needs to have its $\Delta$ updated. This can be done through one tree traversal. When a server is moving, the tree needs to be updated as well. However, the $\Delta$ values are quiescent to server movement and do not need to be updated. The tree is updated as:  
\begin{itemize}
\item Vehicles follow their routes and update the server when a new pickup or dropoff location is reached; Server drops the inactive portion of the tree accordingly;
\item Many moving object indexing methods have been proposed that includes RUM-tree, TRP-tree, Bx-tree, Bdual-tree, and STRIPES. Indexing can substantially decrease the searching of the candidate taxis. However, a trade off needs to be made between maintaining a complex and search-efficient index and relying on a search-approximate but easy to maintain index. In our dataset, around 1,7000 taxis update their locations every 20 to 60 seconds. We choose to use a simple grid-based spatial index. The index is updated when a vehicle moves across boundaries of the index bounding box. For each request, it identifies the vehicles possibly within $w$ of the request, asks the vehicle's actual location, and then tests if these vehicles can accommodate the request.
\end{itemize}

%\subsection{Complexity Analysis}
%Space and time cost... 

%\subsection{Tree Update and Road Network}

%[how the tree is updated when vehicles move]

%[how the tree is updated when an insertion happens]

\section{Hotspot Based Optimization}
\label{large}

The main problem with the basic tree algorithm is the exponential explosion of the size of the tree when there are multiple pickup or dropoff locations close to each other. For example, if we have 8 pickups occur in spatial proximity around similar time. e.g airport terminals, any permutation of the pickups may result in a valid schedule. So there are 8!= 40,320 possibilities already without considering the dropoff points. We propose an approximation approach with bound to reduce the search space. The idea is that when the time and space requirement of computing the best schedule are too much, a server may decide to shed the load by only maintaining a subset of the schedules. Since the number of leaves of the kinetic tree is determined by the number of possible routes, the tree size is effectively controlled by the approximation and the service contraints. %Strategically choosing the subset so that they still allow good ridesharing and competitiveness is challenging.

We propose the following {\bf hotspot clustering algorithm} to deal with this situation. When we insert a pickup point $s_k$ to an edge $(x_i,x_{i+1})$, we check if $d(x_{x+i},s_k) \leq \theta$ where $\theta$ is a small number. If so, $s_k$ is inserted into the node of $x_{i+1}$. $s_k$ and $x_{i+1}$ are treated as one point called {\em hot spot} in the tree and an arbitrary schedule is chosen among the points in a hot spot. When the hot spot contains more than one point, the newly inserted point needs to be within $\theta$ to all the points of the hot spot.   Similar procedure can be done for the dropoff points and the mixture of pickup and dropoffs. Once the point is combined with any node, we stop trying to insert it to any other edges.

\begin{figure}[bt]
\centering
\includegraphics[width =0.5 \textwidth]{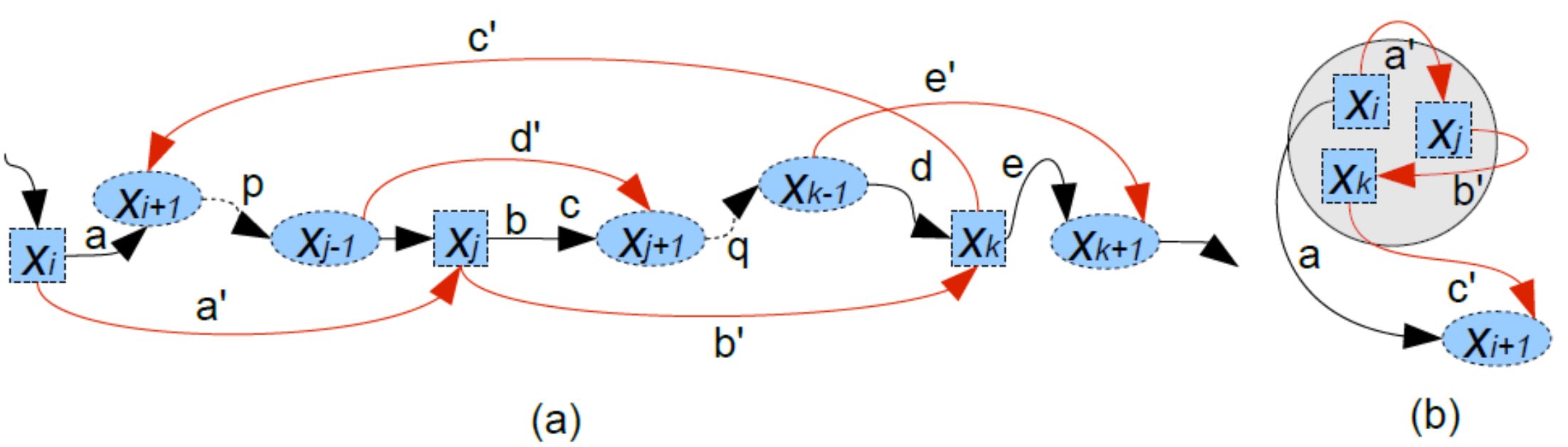}
\caption{Bound for Hotspot. $x_i$, $x_j$, and $x_k$ are in one hotspot.   Black lines: optimal schedule $S_{best}$. We can convert $S_{best}$ by connecting $x_i,x_j, x_k$ consecutively first and then thread the other locations (represented by ovals). The new schedule has a bounded cost.}
\vspace*{-0.4cm}
\label{fig:hotSpot}
\end{figure}

Let us first assume the service constraints are sufficiently large that all schedules are possible.
For a trip set $TR$, let $S_{best}$ be the optimal schedule. Suppose there is a hotspot $hp$ among the pickup and dropoff locations of $TR$. Our hotspot-based method chooses an arbitrary schedule $T_{hs}$ that goes through the points of the hotspot in a consecutive manner. We want to prove that the cost of $T_{hs}$ is bounded.

\bthm 
\label{them:bound}
$cost(T_{hs}) \leq cost(S_{best}) + 2(m+1)\times \theta$ where $m$ is the number of points in the hotspot without considering constraints.
\ethm 

\bproof
We prove when $m=3$ by illustration. For general $m$, the proof is mainly the same.
In Figure \ref{fig:hotSpot} (a), assume $\{x_i,x_j,x_k\}$ has pairwise distance of no greater than $\theta$. The optimal schedule $S_{best}$ is labeled by black solid and dashed lines. We can convert $S_{best}$ into $T_{hs}$ by connecting $x_i,x_j, x_k$ consecutively first and then thread the other locations (represented by ovals) in $S_{best}$ as shown by the red lines and black dashed lines.  We prove that (1) $cost(T_{hs}) \leq cost(S_{best}) + 3\theta$ which is equivalent to prove $a'+b'+c'+d'+e' \leq a+b+c+d+e+3 \times \theta$ since the dashed lines are common in both schedules.

We know $d' \leq b+c$, $e' < d +e$. Now we only  need to show $a'+b'+c' \leq a + 3 \times \theta$. As shown in \ref{fig:hotSpot} (b), we can easily prove that $c' \leq a + \theta$ because the shortest path between $x_k$ and $x_{i+1}$ is no longer than than the schedule $x_k,x_i,x_{i+1}$. Because $b' \leq \theta$ and $c' \leq \theta$,  we know $a'+b'+c' \leq a + 3 \times \theta$.

However, the hotspot algorithm may not use the same order of $x_i,x_j,x_k$ as in the optimal solution as it is an arbitrary order, we now approve that (2) for any two hotspot-based schedule $S_{hs}$ and $S_{hs'}$, $cost(S_{hs}) \leq cost(S_{hs'}) + (m+1) \theta$ where $m=3$. 
Without loss of generality, let $S_{hs} = \ldots, x_{i-1}, x_i,x_j,x_{k}, x_{k+1}\ldots$ and $S_{hs'} = \ldots, x_{i-1}, x_j,x_k,x_{i}, x_{k+1}\ldots$. It is obvious that $d(x_{i-1},x_i)$ $ \leq d(x_{i-1},x_j) + \theta$ and  $d(x_{k},x_{k+1}) \leq d(x_{i},x_{k+1}) + \theta$. Also $d(x_i,x_j) \leq d(x_j,x_k) + \theta$ and $d(x_j,x_k) \leq d(x_k,x_i) + \theta$. Adding the inequalities together, we have $cost(S_{hs}) \leq cost(S_{hs'}) + 4 \theta$
 
Putting (1) and (2) together, we have $cost(S_{hs}) \leq cost(S_{best}) + (2m+1)\times \theta$ where $m=3$.
\eproof

Because after we build the whole tree, we select the shortest schedule with hotspot $cost(S_{hsBest})$ and it is obvious that $cost(S_{hsBest})$ $ \leq cost(S_{best}) + 2(m+1)\times \theta$.

%\begin{comment}
When we consider the constraints,  for $S_{best}$ the corresponding hotspot-based schedule with constraint may violate some constraints and thus does not exist. However, when the constraints of points of the best schedule are relaxed, the corresponding hotspot-based schedule will be found. We have the following theorem.

\bthm 
$cost(S_{hs}) \leq cost(S_{best}) + 2(m+1)\times \theta$ where $m$ is the number of points in the hotspot when constraints of all points in $S_{best}$ is larger than $m \theta$.
\ethm 

\bproof
Again we prove for $m=3$ because of the ease of illustration. In Figure \ref{fig:hotSpot} (a), if $(a,p,b,c,q,d,e)$ is a valid partial schedule with each node having at least $3 \theta$ slack time, then $(a',b',c',p,d',q,e')$ is a valid partial schedule. 

%with each note other than $\{x_i,x_j,x_k\}$  having a slack time of at most $3 \times \theta$ less.

For any point on $p$, the extra delay is $a'+b'+c'-a \leq 3 \times \theta$. For any point on $q$, the extra delay $a'+b'+c' -a + d' - (b+c ) \leq a'+b'+c' \leq 3 \times \theta$. For $x_{i+1}$, the extra delay is $a'+b'+c'+d'+p+q - (a+b+c+d+p+q)$ which is proven in Theorem \ref{them:bound} as no larger than $3 \times \theta$.
\eproof

When $\theta$ is sufficiently small, we will likely to find a schedule that is upper bounded by the best schedule with a small additional time. %The alternative is to identify a central location for each hotspot and allow people to walk to the central location for pickup and dropoff.

%\section{Nearest Neighbor Search and Fast Pruning Approach}
%\label{fast}

%In the current approach, for a given request $tr_i$, we first find all the taxi with $d_T(r,s_i) \leq R_i$. 
%To do that, we can perform a range query to find out all the taxi with range $d(r,s_i) \leq R_i$. 
%Then, for each taxi, we will traverse its corresponding prefix to find out which edge can accommodate a detour to $s_i$. 
%A major issue here is the insertion of the pickup location and the dropoff location considered independently. 
%Thus, even the taxi can pickup the customer, but the dropoff would violate detour constraints. 
%Can we quickly prune some insertions of the pickup location which cannot handle the dropoff location?

%We utilize the following result: 

%\blemma
%For any candidate edge $(x_j,x_j^\prime)$ to handle trip request $tr_i$, the following condition must hold: 
%$d(r,x_j)+d(x_j,s_i) \leq R$ and $d(s_i,x_j^\prime)+d(x_j^\prime,e_i) \leq (1+\epsilon) d(s_i,e_i)$. 
%\elemma

%Given this, we can perform a Dijkastra's algorithm for $s_i$ and search for all $x_j$ and $x_j^\prime$ which satisfy this constraint with range $R$. 
%Basically, we can quickly discover the candidate edges to insert $s_i$ without performing the BFS or DFS seach of the tree of each taxi with the range of $R$. 
%We may add additional information in the road network to facilitate such search. 
%Note that once we find the candidate edge, we still need to apply condition (a) and (b) to do the final test. 

%[Add the reverse Dijkastra algoritm from $s_i$ in searching candidate edge pair $(x_j,x_j^\prime)$.]

%\end{comment}

\begin{figure*}[ht!]
\begin{center}
	\subfigure{\scalebox{0.22}{\includegraphics{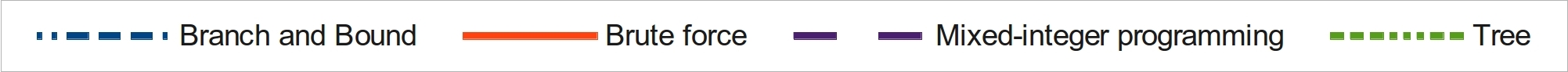}}} \\
	\subfigure{\scalebox{0.15}{\includegraphics{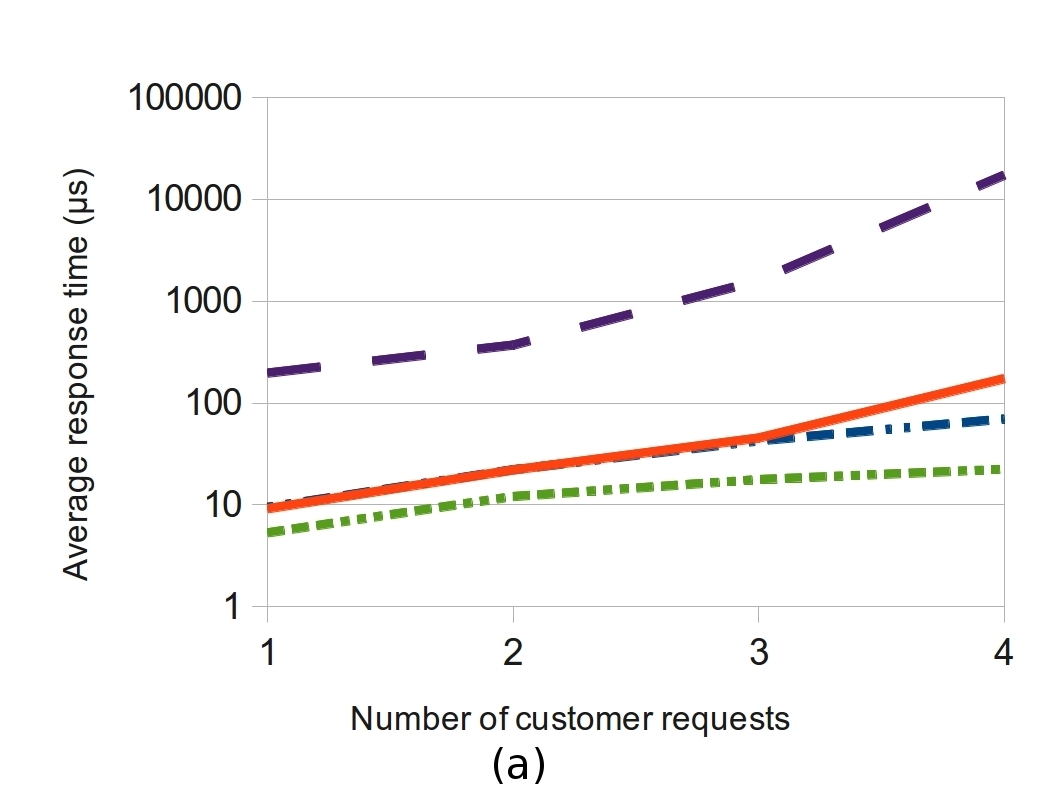}}}
	\subfigure{\scalebox{0.15}{\includegraphics{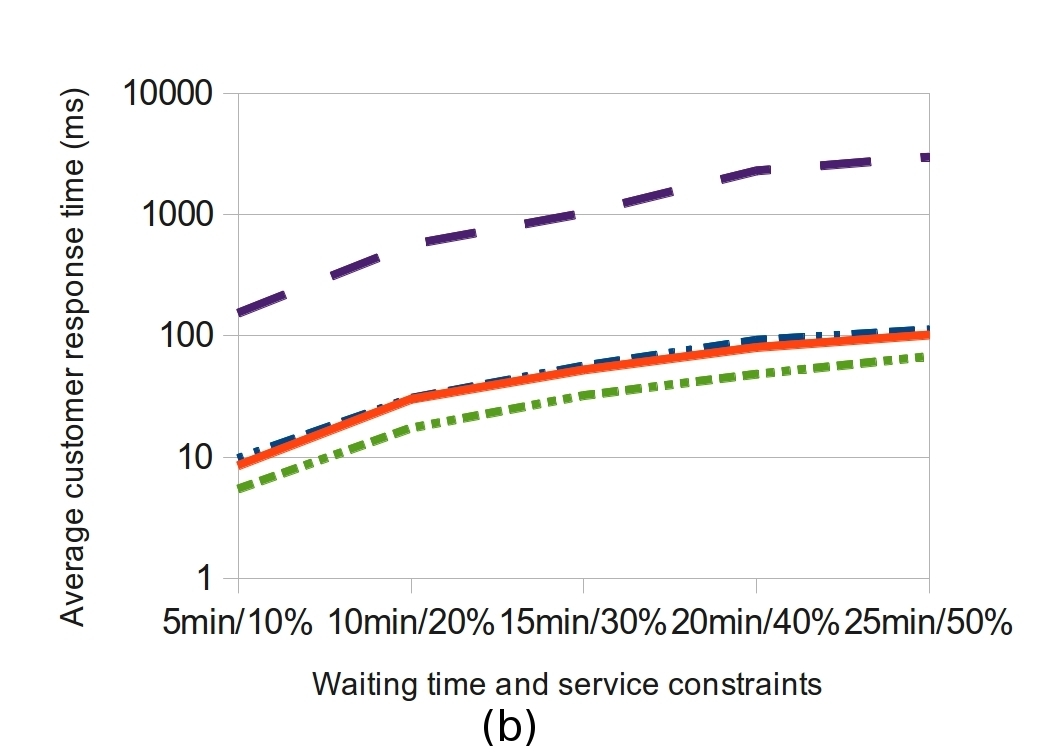}}}
	\subfigure{\scalebox{0.15}{\includegraphics{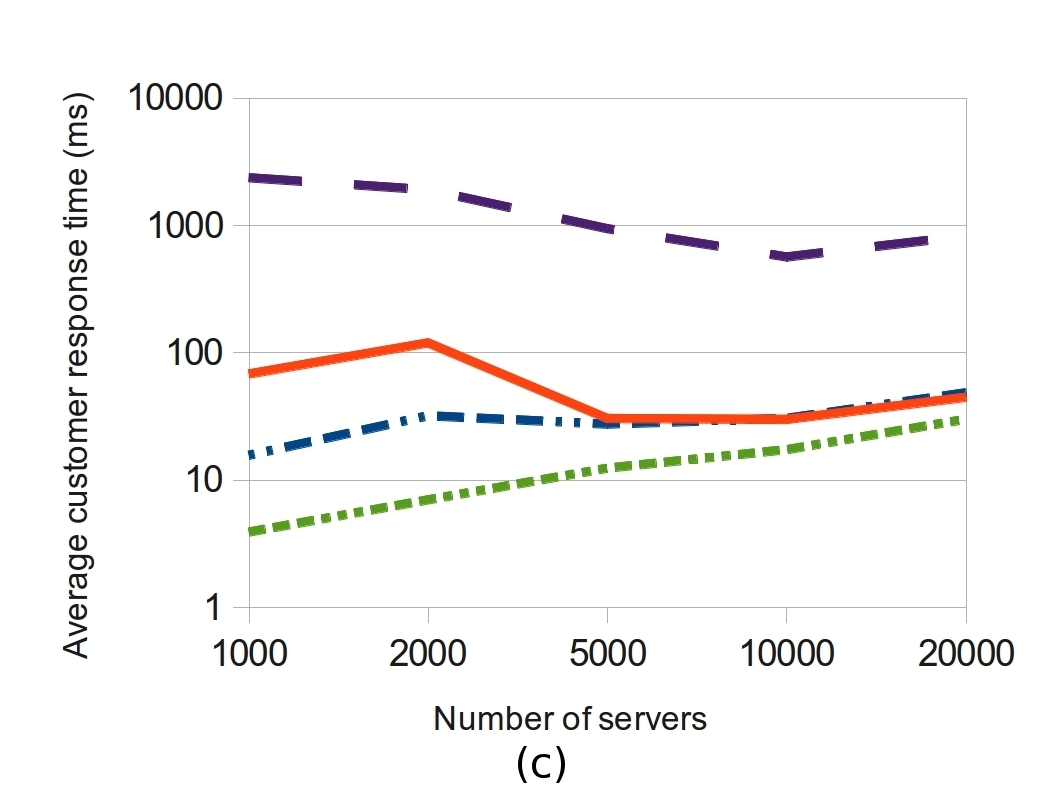}}} 
	
	\caption{Results for changing number of customer requests (a), constraints (b), and number of servers (c). Default parameters are 10 min / 20\% for the constraints, 10,000 servers, and a capacity of 4.}
	\label{fig:expr_result1}
\end{center}
\end{figure*}

\section{Experimental Design} \label{experiment}
Data for the experiments is based on trips of 17,000 Shanghai taxis for one day (May 29, 2009); the dataset contains 432,327 trips. Each trip $t$ includes the starting and destination coordinates $t.s$ and $t.e$ and the start time $t.time$ (i.e., the time at which the taxi picked up the passengers for the trip).

%This data is used for all algorithms and in all test cases, meaning all results should be the same (except for the tree clustering algorithm, which could have some loss of optimality). 

A simulation framework submits trip requests to the system in real-time based on the trips. Specifically, for each trip $t$, and trip request is initialized as $tr = <s_i = t.s, e_i = t.e>$, and $tr$ is submitted at time $t.time$ in the simulator.

An instance of the route algorithm is associated with a vehicle. When a new trip request is received, the simulator trips the request with each vehicle and then chooses the vehicle returning the minimum time; the request is then assigned to that vehicle.

The simulation framework executes based on an undirected, weighted graph derived from and representing the Shanghai road network. This graph contains 122,319 vertices and 188,426 edges. The starting and destination trip coordinates are pre-mapped to the closest vertex in the graph. This may result in some inaccuracy if the coordinates are in the middle of a street or if the coordinates do not match with the road network data, but this inaccuracy is negligible. A vehicle is initialized to a random vertex in the city, and then follows a given route when there are customer(s) on board or, otherwise, follows the current road segment (at intersections, the next segment to follow is chosen randomly). We assume that speed in the road network is a constant 14 meters/second (approximately 48 kilometers/hour). Then, most computations are done in terms of distance instead of time (such as finding the path with shortest duration).

The framework is implemented in C++. We run the experiments on cluster nodes with an Intel Xeon X5550 (2.67GHz) processor. The simluation implementation is single-threaded, so only one core of the CPU is used. We limit the memory usage of each simulation process to three gigabytes.

%todo: add how to calculate the shortest path!
To simulate the taxis, we need both the distance and routes (vehicles follow) between vertices in the road network. Computing shortest path on road networks has been widely studied (see ~\cite{Delling:2009} for an extensive review).
A variety of techniques~\cite{Delling:2009}, such as $A^\ast$,  Arc-flag (directing the search towards the goal), highway hierarchies (building shortcuts to reduce search space, transit node routing (using a small set of vertices to relay the shortest path computation), and utilizing spatial data structures to aggressively compress the distance matrix, have been developed.
Recently, Abraham {\em et al.}~\cite{Abraham:2010:HDS} recently discovered that several of the fastest distance computation algorithms need the underlying graphs to have small {\em highway dimension}. Furthermore, they demonstrate the method with the best time bounds is actually a labeling algorithm~\cite{Abraham:2010:HDS}. We implement the state-of-art hub-labeling algorithm - a fast and practical algorithm to heuristically construct the distance labeling on large road networks, where each vertex records a set of intermediate vertices (and their distance to them) for the shortest path computation~\cite{Abraham:2011:HLA}. For the purposes of tracking taxi location, a second version of the road network is stored in memory in a weighted adjacency list structure without additional information.

However for large scale ridesharing, the shortest path algorithm is called very frequently and can be the bottleneck if not implemented efficiently. We observe the repeated calling follows a pattern that preserves locality. So, we implement two {\bf LRU caches} using a single hash table, one storing up to ten million shortest distances and the other storing up to ten thousand shortest paths (separate caches are used because more distances can be stored in memory, and shortest distance is needed more often than shortest path). Both caches are indexed only by the starting and destination points in a distance or path computation call; this is accomplished by defining the index for two vertices $s$ and $e$ as $i = \textrm{id}(s) \cdot |V| + \textrm{id}(e)$, where id returns an integer representation for a vertex.

\begin{table}[h]
\begin{tabular}{|c|c|c|}
\hline
Parameter & Tested settings \\ \hline
\hline
Capacity & {\bf 4} \\
\hline
Constraints & 5 min / 10\%; {\bf 10 min / 20\%}; \\
& 15 min / 30\%; 20 min / 40\%; 25 min / 50\% \\
\hline
Number of servers & 1,000; 2,000; \\
& 5,000; {\bf 10,000}; 20,000 \\
\hline
\end{tabular}
\caption{Parameters for four-algorithm comparison.}
\label{expr_param}
%\vspace{-1cm}
\end{table}

\subsection{Four Algorithm Comparison}
We first compare kinetic tree algorithm with the branch and bound, brute-force, and mixed-integer programming algorithms under the dataset of 432,327 trips.

We choose three important parameters: capacity, waiting time and tolerance constraints, and number of taxis/servers. We first establish reasonable defaults for the parameters, and then proceed to modify the parameters one at a time to evaluate their effect. The defaults (bolded) and other tested settings are shown in Table \ref{expr_param}. Note that a waiting time constraint of 10 minutes corresponds to 8,500 meters.

To evaluate performance, we measure the average customer response time (ACRT), the average time required to complete the search for the minimum time needed to satisfy a new request.  We further measure the average response time (ART), the average time needed to calculate the best route for a taxi to follow given its current state, for different request sizes. Depending on the number of  requests need to be scheduled, the ART can change significantly (for example, a taxi with twenty current requests would have forty more points to schedule than one with no assigned requests). Thus, we calculate ART separately for different current request sizes and then compare to see the effect of the number of current requests on response time.

%Because the framework runs through the one-day simulation period, we measure both the overall ACRT and the ACRT for the last thousand requests.

%This is the average time needed to calculate the best route for a taxi to follow given its current position, its current requests, and new requests.
 
The default taxi capacity is set at four both to mimic real-world situations and because a higher capacity results in other algorithms not being able to finish executing within a reasonable time. Additionally, rather than testing lower capacities, we use ART to find the effect of different problem sizes on the efficiency of the algorithms.

Figure \ref{fig:expr_result1} shows the four-algorithm comparison. Figure \ref{fig:expr_result1} (a) shows the ART with different numbers of requests. Figures \ref{fig:expr_result1} (b) and (c) show the ACRT for varying constraints and fleet size, respectively. Generally, the brute-force and branch and bound algorithms exhibit roughly the same performance. The mixed-integer programming approach takes significantly more time, probably because of significant execution time used to initialize and preprocess each mixed-integer programming problem. The tree algorithm outperforms the other algorithms for all test cases, due to its incremental approach.

%the axis labels include the new request (so 2 means that there is one request already assigned to the taxi, and another is the new request just being assigned).

For a small number of taxis and a large number of customers already scheduled to the taxi, branch and bound outperforms brute-force. The reason is most likely that the pruning effect of branch and bound is more important when the shortest route calculations have more customer requests, because the problem size is larger. When the problem size is small, the fast initialization of the brute-force algorithm is preferable; branch and bound, on the other hand, has to first calculate the minimum edges for each of the vertices in the complete graph of pickup and dropoff points that it uses.

\begin{figure*}[t]
\begin{center}
	\subfigure{\scalebox{0.22}{\includegraphics{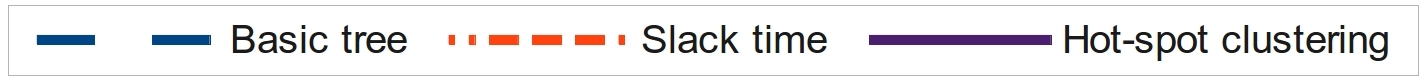}}} \\
	\subfigure{\scalebox{0.15}{\includegraphics{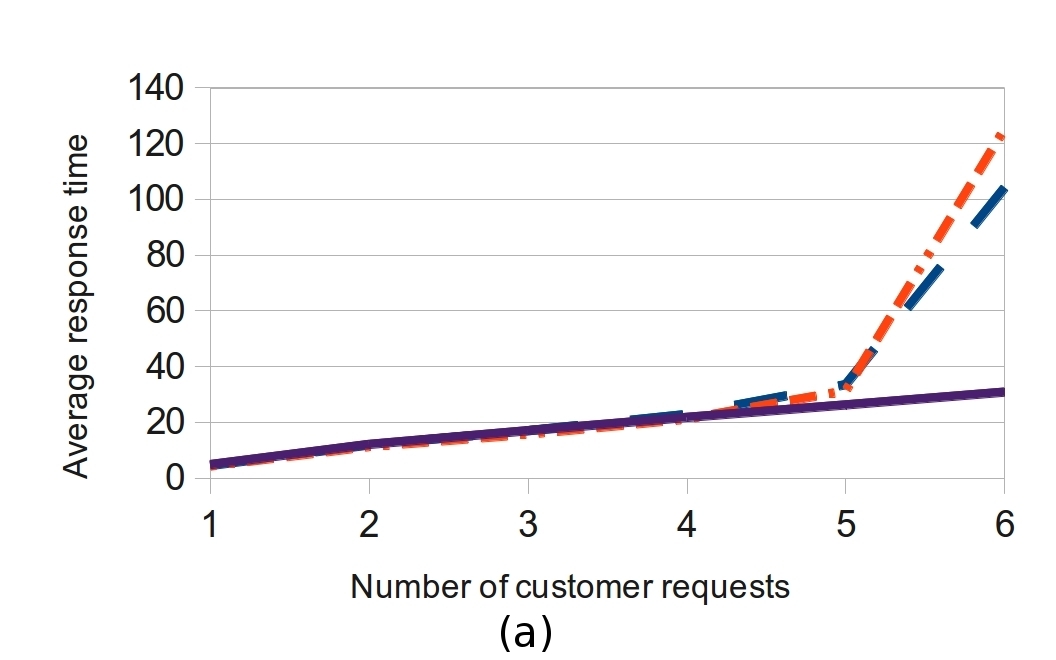}}}
	\subfigure{\scalebox{0.15}{\includegraphics{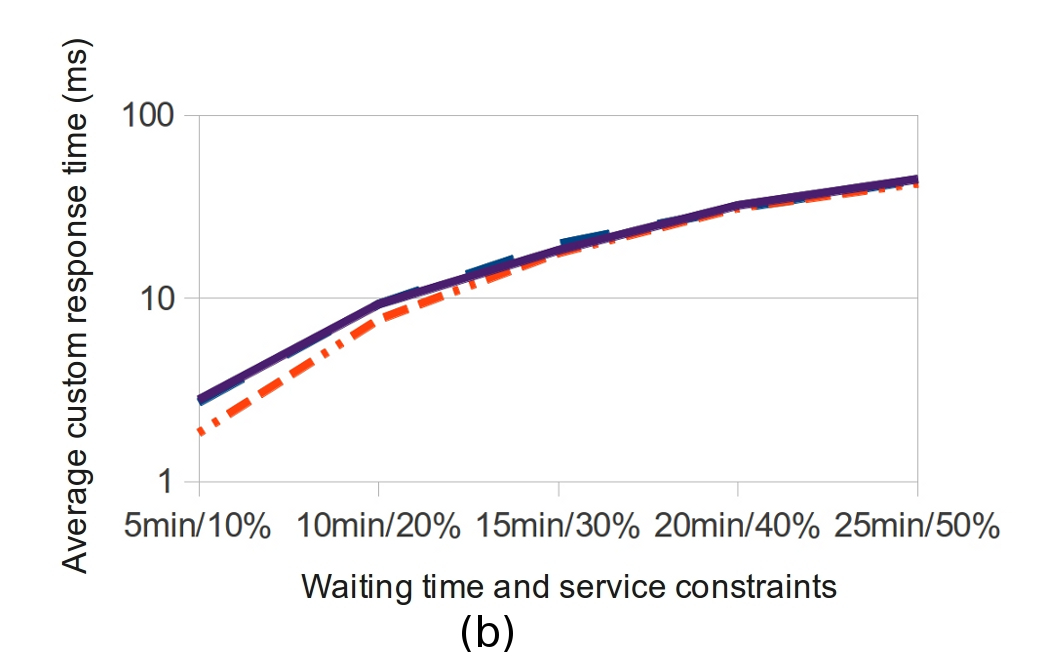}}}
	\subfigure{\scalebox{0.15}{\includegraphics{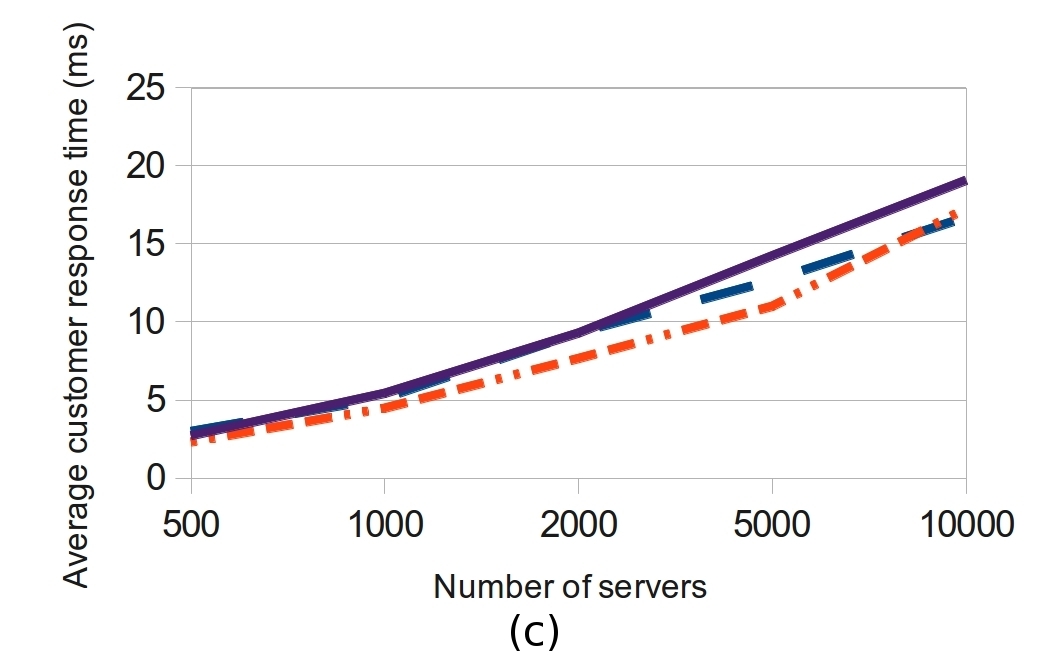}}} 
	
	\caption{Results with tree algorithms for different numbers of customer requests (a), changing constraints (b), and changing number of servers (c). Default parameters are 10 min / 20\% for the constraints, 2,000 servers, and a capacity of 6 ((a) shows ART in different cases for these parameters).}
	\label{fig:expr_result2}
\end{center}
\end{figure*}

For the default parameters, the execution time of the branch-and-bound and brute-force algorithms are almost identical, while the mixed-integer programming is approximately 20 times slower. The tree algorithm, on the other hand, is almost two times faster than the branch-and-bound algorithm. Similar magnitude execution time differences are seen for other parameters.% with the tree algorithm leading the mixed-integer lagging in terms of speed.

\begin{figure}[h]
\begin{center}
	\subfigure{\scalebox{0.2}{\includegraphics{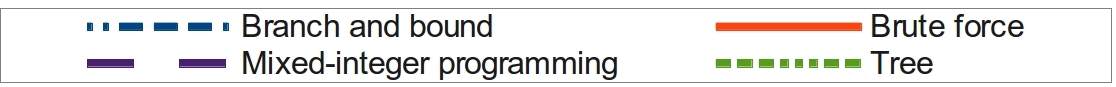}}}
	\subfigure{\scalebox{0.2}{\includegraphics{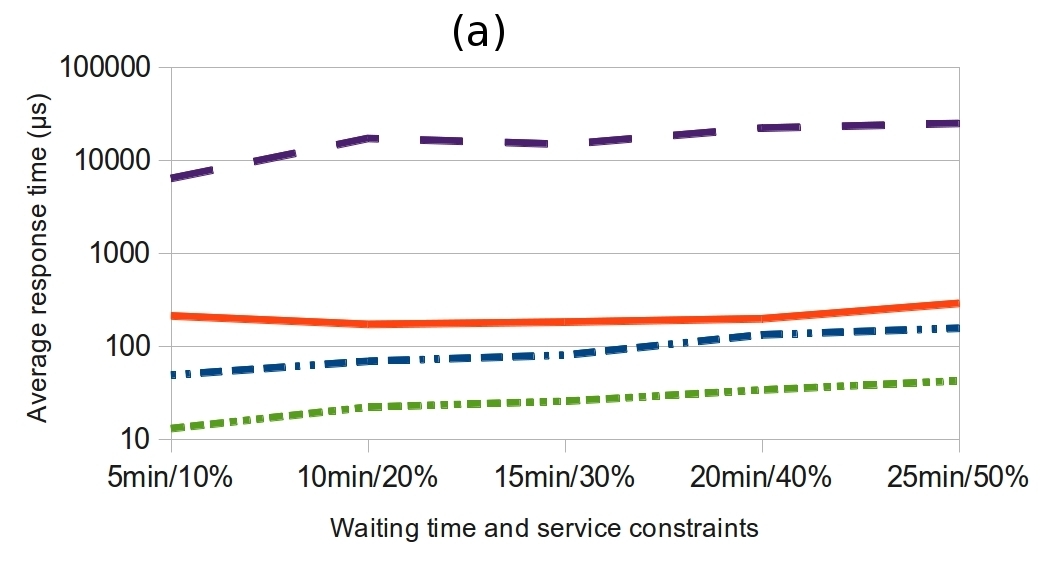}}}
	\subfigure{\scalebox{0.2}{\includegraphics{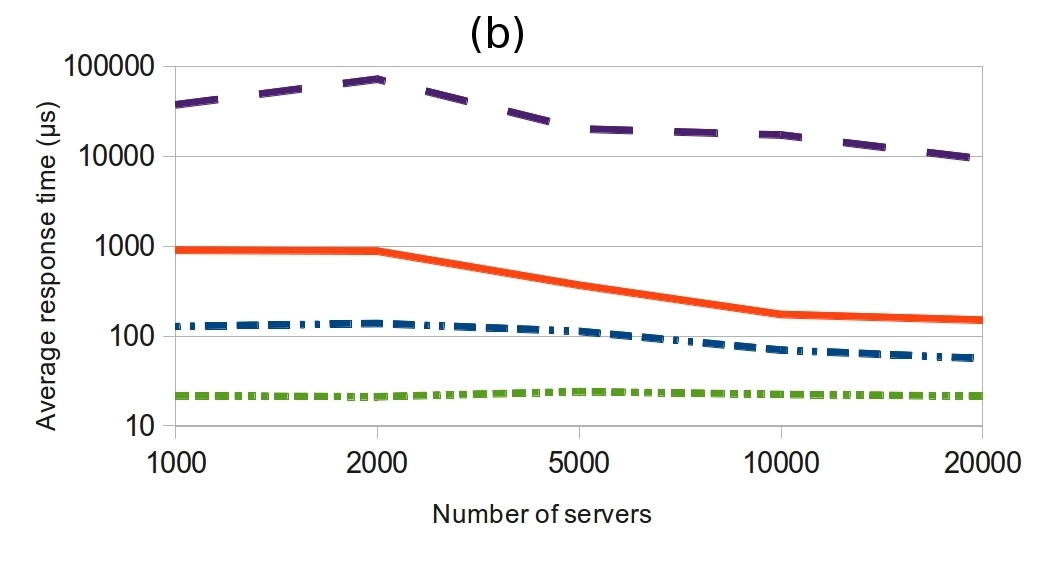}}}
	\caption{ART for four customer requests when changing constraints (a) and number of servers (b). Default parameters are 10 min / 20\% for the constraints, 10,000 servers, and a capacity of 4.}
	\label{fig:expr_result1_mc}
\end{center}
\end{figure}

In Figure \ref{fig:expr_result1_mc}, we show ART when shortest routes are being calculated for four customer requests to get an idea of performance of larger sizes. These are graphed as constraints or number of servers increase. For constraints, the ART gradually increases as the constraints become looser. This makes sense because more feasible combinations that have to be considered. Additionally, the brute-force appears to be less influenced by this trend, probably because the brute-force anyway enumerates each permutation(the constraints still does effect its ART because it can stop earlier on average when checking the feasibility of each permutation).

The increasing number of servers appears to have little effect on the tree algorithm, but for the other three algorithms, the ART clearly decreases. This is most likely because with more servers, most of the cases where there are four passengers occur when the pickup and dropoff points of the passengers are not as clustered: there will often be an empty server close to a server with several passengers, and it would get the next request. Because the pickup and dropoff points are farther apart, there are less combinations and the execution time is lower. On the other hand, when there are few servers, the servers are spread farther apart, meaning that it is more likely for a single server to handle several clustered requests.

\subsection{Comparing Tree Algorithms}
We further evaluate different versions of our tree algorithm: basic tree algorithm, the slack time algorithm, and the hot-spot clustering algorithm (which also uses slack time).

\begin{table}[h]
\begin{tabular}{|c|c|c|}
\hline
Parameter & Tested settings \\ \hline
\hline
Capacity & 3; 4; 5; {\bf 6}; 7; 8; 12; 16; unlimited \\
\hline
Number of servers & 500; 1000; {\bf 2000}; 5,000; 10,000 \\
\hline
Constraints & 5 min / 10\%; {\bf 10 min / 20\%}; \\
& 15 min / 30\%; 20 min / 40\%; 25 min / 50\% \\
\hline
\end{tabular}
\caption{Parameters for Tree algorithm Comparison.}
\label{expr_treeparam}
\vspace{-0.5cm}
\end{table}
We set a default capacity of six because we find that the tree algorithms are able to solve larger problems than the other approaches. %Furthermore, because the tree algorithms are very similar, we measure the ACRT after varying the capacity of the taxi/server in addition to measuring ART for the default capacity of six.
The incremental nature of the tree algorithms and improvements in the hot-spot clustering algorithm allows us to explore the effect of an unlimited capacity (for most of the other algorithms, we find that this leads to too great a search space), which gives an idea of what the maximum achievable ridesharing is. The parameters we use for evaluating the tree algorithms are shown in Table \ref{expr_treeparam}, with the bolded values being the default settings. %We keep two parameters constant at the default settings and then vary either the constraints, the capacity, or the number of servers.

\begin{figure*}[htbp]
\begin{center}
	\subfigure{\scalebox{0.22}{\includegraphics{figures/e2Legend}}} \\
	\subfigure{\scalebox{0.15}{\includegraphics{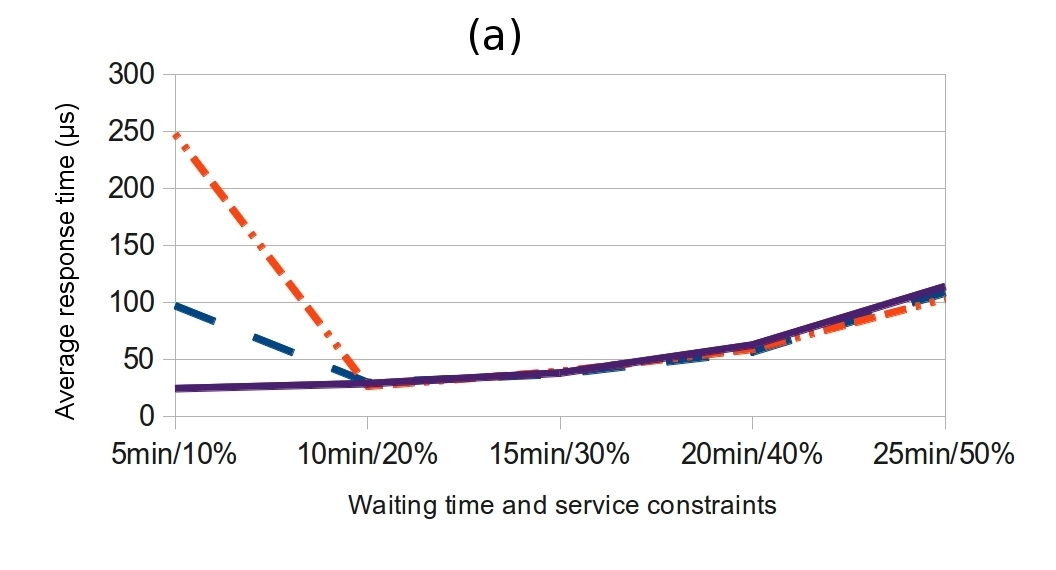}}}
	\subfigure{\scalebox{0.15}{\includegraphics{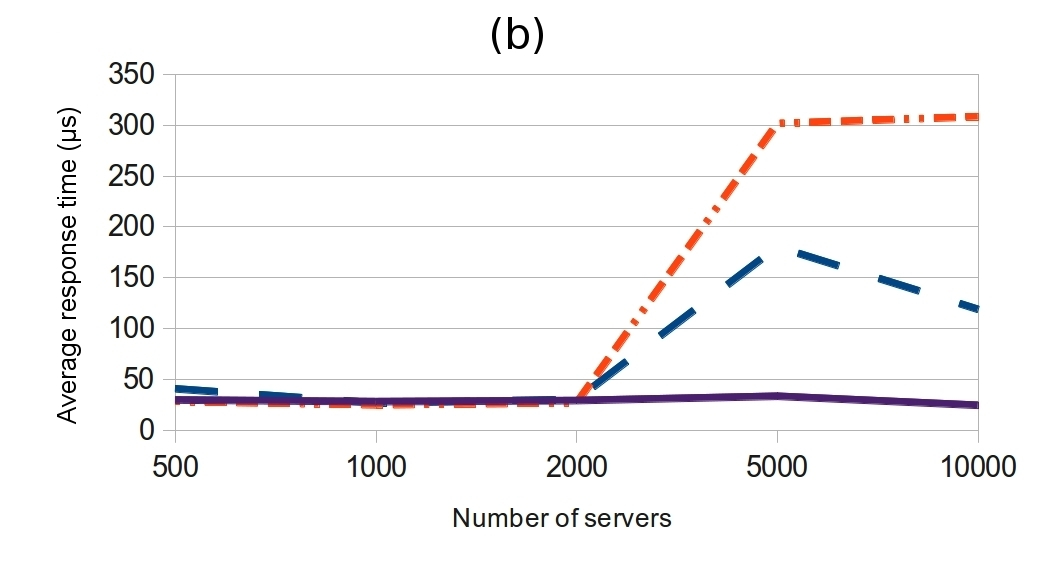}}}
	\subfigure{\scalebox{0.15}{\includegraphics{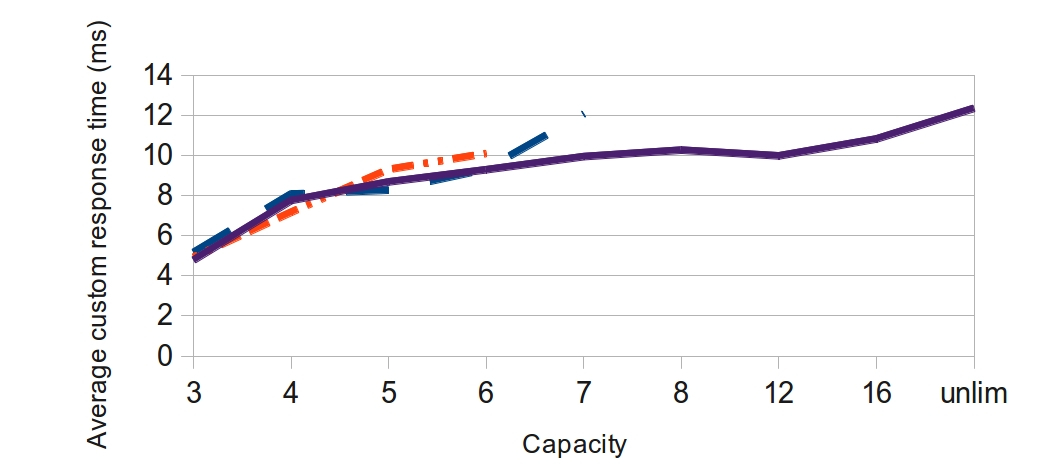}}}
	
	\caption{ART for six customer requests when changing constraints in (a) and number of servers in (b). In (c), tree-algorithm ACRT for different capacities. ``unlim'' indicates unlimited capacity. Only hotspot clustering algorithm can complete for unlimited capacity. Default parameters are and a capacity of 6, 10 min / 20\% for the constraints and 2,000 servers. }
	\label{fig:expr_result2_mc}
\end{center}
\end{figure*}

We now evaluate the performance of the slack time and hot-spot clustering improvements that we make to the tree algorithm. Figure \ref{fig:expr_result2} shows these results. The slack-time algorithm is faster than the basic tree algorithm except when the number of servers is 10,000, the number of customer requests is 6, or the constraints are at 20 min / 40\%; these cases are examined more closely in Figure \ref{fig:expr_result2_mc}. Slack-time achieves a maximum time saving of over 32\% compared to the basic tree algorithm when the constraints are at the tighest level tested, 5 min / 10\%. For the default parameters, it yields savings of approximately 18\%. So, the slack-time does yield a relatively significant improvement in time, especially when constraints are tight so that many branches in the tree are infeasible.

Figure \ref{fig:expr_result2_mc} presents results similar to those in Figure \ref{fig:expr_result1_mc}, but for the tree algorithms. Most prominent in the graphs is the steep increase in ART for tight constraints and large capacities with the basic and slack-time tree algorithms; this is opposite to the results in Figure \ref{fig:expr_result1_mc}. It can be explained, however, by the increased capacity that we use. In both cases where the ART is large, it is relatively rare for a server to have six passengers: typically, there would either be another server with less passengers available to handle the request or the constraints would be too tight to allow so many passengers. So, when the server is able to get six passengers, it is most likely because the pickup/dropff points are very close to each other. In these cases, the short distance between the points creates a large number of feasible combinations. Although these cases would also appear for looser constraints and smaller numbers of servers, the ART is an average, and other six-passenger-cases that do not create a large number of combinations would be much more common. This also explains why the hot-spot clustering algorithm is not affected by the trend. The slack-time tree algorithm is faster than the basic tree algorithm because slack time only reduces execution time when there are many infeasible branches that can be pruned.

Additionally, like in Figure \ref{fig:expr_result1_mc}, the ART in Figure \ref{fig:expr_result2_mc} increases gradually with looser constraints starting from the 10 min / 20\% parameters because looser constraints also allow more feasible combinations.

%\begin{figure}[htbp]
%\begin{center}
%	\subfigure{\scalebox{0.2}{\includegraphics{figures/e2LegendMC}}}
%	\subfigure{\scalebox{0.2}{\includegraphics{figures/e2Capacity}}}	
%	\caption{Tree-algorithm ACRT for different capacities. ``unlim'' indicates unlimited capacity. Default parameters are 10 min / 20\% for the constraints and 2,000 servers. Only hotspot clustering algorithm can complete for capacity upto ``unlim''.}
%	\label{fig:expr_result2_capacity}
%\end{center}
%\end{figure}

Figure \ref{fig:expr_result2_mc} (c) shows the ACRT results for different capacities. The ACRT breaks off for each algorithm when it can no longer finish in a reasonable time or exceeds the imposed memory limit of three gigabytes. The hot-spot clustering algorithm is the only one that is able to finish the simulation program with a capacity greater than seven, and also for unlimited capacity (marked as unlim in the figure).

From this figure, we can see that while the basic and slack-time tree algorithms are unable to continue processing when the problem sizes become too large, the hot-spot clustering algorithm is scalable to higher capacities. This also confirms our hypothesis that the biggest issue for unlimited capacity is situations where a large number of passengers wish to depart from a single point; hot-spot clustering combines these points in the tree.

The maximum number of passengers at unlimited capacity in a single server is 17, while the average is 1.7 (this is with the default parameters, so the number of servers is two thousand). The average in the top 20\% filled servers is a bit higher than 3.9. This indicates that the majority of vehicles in a server fleet should be five-person cars (with one of the five seats taken by the driver), but for some requests larger vehicles are needed.

\section{Related Work} 
\label{relate}
Our work is related to nearest neighbor (NN) search on moving objects over road networks.
Early work on nearest neighbor search on road networks focuses on data models that are easy to implement and
serve as a foundation for NN queries \cite{Jensen:2003}. Later research has focused on continuous monitoring of
nearest neighbors (NN) in highly dynamic scenarios, where
the queries and the data objects move frequently on a road network \cite{Mouratidis:2006}.
A recent paper addresses the problem of monitoring the $k$ nearest
neighbors to a dynamically changing path in road networks. Given a destination where a user is going to, this
new query returns the k-NN with respect to the shortest
path connecting the destination and the user's current location \cite{Chen:2009:MPN}.
Guting {\em et. al.} proposed algorithms to find the $k$ nearest neighbors to $m_q$ within $D$ for any instant of time within the lifetime of $m_q$ given a set of moving object trajectories $D$ and a query trajectory $m_q$ \cite{Guting:2010}. Nearest neighbor query on
road network is only the minor step in the ridesharing system that can help to filter the initial set of candidate taxis.

The {\em trip grouping} algorithm \cite{Gidofalvi:2008} groups ``closeby'' cab requests using a set of heuristics. Requests are queued for a user given waiting time to be scheduled.  The heuristics include grouping requests upon expiration, estimation combination saving using pairwise request combination gain, and greedy grouping. The trip grouping algorithm is then expressed as a continuous stream query and optimized by space partitioning and parallelization. This method is heuristic-based and does not provide  waiting and riding time service guarantee as our method does.

In operation research, early research on this problem mostly focuses on a single vehicle and a static scenario where the set of requests are known ahead of time. This is unrealistic for large scale and ad-hoc services such as a taxi service. The problem is, unsurprisingly, NP-hard. Only problems with small sizes can be solved to optimality.  Exact dynamic programming algorithms have been developed \cite{Psaraftis:1983}.  Note that  the problem without a deadline can be considered as the special case of the problem with a deadline where the deadline is infinite. Once the fixed deadline is given, we can construct subproblems using these deadlines, and thus dynamic programming can be employed. However, in our case, since the maximal waiting time and the service level are two separate constraints, each trip request can be enforced with a fixed completion deadline. Thus, the dynamic programming approaches can not be applied to our problem. We also note that our problem can be considered more general than the fixed deadline problem. Given a fixed deadline $t$, the maximal waiting time can be defined as $w=t-(1+\epsilon)d(s,e)$. Thus, our algorithm can also be used for the fixed deadline problem.

In a dynamic single vehicle DARP problem, requests come in real time and a server has to make decisions on-line \cite{Feuerstein:2001}.  In the problems without deadline, the objectives are to minimize {\em makespan} (time to finish the last request) or the {\em average completion time}. {\em Competitive ratio} is a standard tool to measure the effectiveness of a dynamic DARP algorithm. An on-line algorithm $A$ is called $c-competitive$ if for any instance $\delta$, the cost of $A$ on $\delta$ is at most $c$ times the offline optimum on $\delta$. This is assuming an optimal solution is available which is false for modern large scale scheduling problem we are addressing . %The costs are typically {\em makespan} and {\em average completion time} without deadline; {\em number of requests served} without deadlines. For dynamic cases without deadlines, a  3.42 competitive ratio in {{\em makespan} algorithm  and a 5.83 competitive ratio in {\em average completion time} algorithm were found.  It is not clear how these algorithms can be extended for same problem with deadlines.

This paper deals with the multiple servers, dynamic (i.e. real-time) DARP  with deadlines. When deadline is involved, the objectives are three folds: (1) real-time response; (2) minimize average completion time; (3) maximize requests served. The most related work is the two-phase insertion technique \cite{Coslovich:2006}.

The single vehicle problems are typically solved to optimality by a branch-and-bound algorithm which may incur exponential time complexity \cite{Colorni:2001}. The state-of-the-art Branch-and-cut ({\em BaC}) algorithm \cite{Cordeau:2006} formulates the multiple server version of this problem using mixed-integer programing and a branch-and-cut solution. {\em BaC} can find exact solutions for small to medium size instances (4 vehicle and 32 requests on a moderate PC for tens to hundreds of minutes). It assumes all vehicles and requests are available ahead of time which is not realistic for a dynamic taxi service of thousands of vehicles serving through out the day. Nevertheless, the solution can be adopted to accommodate the attempts of combining new requests with existing routes of vehicles. We compare our kinetic tree based approach to a branch-and-bound approach and a mixed integer programing approach in this paper.

%The DARP problems can be static or dynamic, with or without deadlines, and  use single or multiple servers. 

%\begin{itemize}
%\item Static single vehicle DARP
%\item Static multiple vehicle DARP \cite{Cordeau:2006}
%\item Dynamic single vechile DARP \cite{Feuerstein:2001,Ascheuer:2000,Lipmann:2002}
%\item Dynamic multiple vehicle DARP \cite{Colorni:2006}
%\end{itemize}

\section{Conclusion}
\label{conclusion}
In this paper, we formulate and propose a kinetic tree algorithm with optimizations to dynamically match real-time trip requests to servers in a road network to allow ridesharing. The proposed algorithm outperforms commonly used approaches including branch and bound and mixed-integer programing, as shown by expriments on a large taxi dataset. In the future, we would like to consider uncertainty issues in scheduling; this is very important and may be a major road block in achieving large scale ridesharing.

\bibliographystyle{IEEETran}
\bibliography{rideshare}

\begin{thebibliography}{10}
\providecommand{\url}[1]{#1}
\csname url@rmstyle\endcsname
\providecommand{\newblock}{\relax}
\providecommand{\bibinfo}[2]{#2}
\providecommand\BIBentrySTDinterwordspacing{\spaceskip=0pt\relax}
\providecommand\BIBentryALTinterwordstretchfactor{4}
\providecommand\BIBentryALTinterwordspacing{\spaceskip=\fontdimen2\font plus
\BIBentryALTinterwordstretchfactor\fontdimen3\font minus
  \fontdimen4\font\relax}
\providecommand\BIBforeignlanguage[2]{{%
\expandafter\ifx\csname l@#1\endcsname\relax
\typeout{** WARNING: IEEEtran.bst: No hyphenation pattern has been}%
\typeout{** loaded for the language `#1'. Using the pattern for}%
\typeout{** the default language instead.}%
\else
\language=\csname l@#1\endcsname
\fi
#2}}

\bibitem{Ghoseiri:2011}
K.~Ghoseiri, A.~Haghani, and M.~Hamedi, ``Real-time rideshare matching
  problem,'' \emph{Final Report of UMD-2009-05, U.S. Department of
  Transportation}, 2011.

\bibitem{Billerburg:2011}
J.~F. Dillenburg, O.~Wolfson, and P.~C. Nelson, ``The intelligent travel
  assistant,'' in \emph{The IEEE 5th International Conference on Intelligent
  Transportation Systems}, 2002, pp. 691--696.

\bibitem{Tickengo:2011}
TICKENGO, ``Tickengo,'' http://tickengo.com.

\bibitem{Stach:2011}
C.~Stach and A.~Brodt, ``vhike - a dynamic ride-sharing service for
  smartphones,'' in \emph{Mobile Data Management}, 2011, pp. 333--336.

\bibitem{Gidofalvi:2008}
G.~Gidofalvi, T.~B. Pedersen, T.~Risch, and E.~Zeitler, ``Highly scalable trip
  grouping for large-scale collective transportation systems,'' in
  \emph{Proceedings of the 11th international conference on Extending database
  technology: Advances in database technology}, ser. EDBT '08, 2008, pp.
  678--689.

\bibitem{Feuerstein:2001}
E.~Feuerstein and L.~Stougie, ``On-line single-server dial-a-ride problems,''
  \emph{Theor. Comput. Sci.}, vol. 268, no.~1, pp. 91--105, Oct. 2001.

\bibitem{Colorni:2001}
A.~Colorni and G.~Righini, ``Modeling and optimizing dynamic dial-a-ride
  problems,'' \emph{International Transactions in Operational Research},
  vol.~8, no.~2, pp. 156--166, 2001.

\bibitem{Cordeau:2006}
J.-F. Cordeau, ``A branch-and-cut algorithm for the dial-a-ride problem,''
  \emph{Oper. Res.}, vol.~54, no.~3, pp. 573--586, 2006.

\bibitem{Tao:2003}
Y.~Tao, D.~Papadias, and J.~Sun, ``The tpr*-tree: an optimized spatio-temporal
  access method for predictive queries,'' in \emph{Proceedings of the 29th
  international conference on Very large data bases}, 2003, pp. 790--801.

\bibitem{Mokbel:2004}
M.~F. Mokbel, X.~Xiong, and W.~G. Aref, ``Sina: scalable incremental processing
  of continuous queries in spatio-temporal databases,'' in \emph{Proceedings of
  the 2004 ACM SIGMOD international conference on Management of data}, 2004,
  pp. 623--634.

\bibitem{Jensen:2004}
C.~S. Jensen, D.~Lin, and B.~C. Ooi, ``Query and update efficient b+-tree based
  indexing of moving objects,'' in \emph{Proceedings of the Thirtieth
  international conference on Very large data bases}, 2004, pp. 768--779.

\bibitem{Kalantari:1985}
B.~Kalantari, A.~V. Hill, and S.~R. Arora, ``An algorithm for the traveling
  salesman problem with pickup and delivery customers,'' \emph{European Journal
  of Operational Research}, vol.~22, no.~3, pp. 377--386, 1985.

\bibitem{Desrochers:1991}
M.~Desrochers and G.~Laporte, ``Improvements and extensions to the
  miller-tucker-zemlin subtour elimination constraints.'' \emph{Operations
  Research Letters}, vol.~36, no.~10, pp. 27--36, 1991.

\bibitem{Delling:2009}
D.~Delling, P.~Sanders, D.~Schultes, and D.~Wagner, ``Algorithmics of large and
  complex networks,'' D.~Lerner, J.and~Wagner and K.~Zweig, Eds., 2009, ch.
  Engineering Route Planning Algorithms.

\bibitem{Abraham:2010:HDS}
I.~Abraham, A.~Fiat, A.~V. Goldberg, and R.~F. Werneck, ``Highway dimension,
  shortest paths, and provably efficient algorithms,'' in \emph{SODA '10},
  2010.

\bibitem{Abraham:2011:HLA}
I.~Abraham, D.~Delling, A.~V. Goldberg, and R.~F. Werneck, ``A hub-based
  labeling algorithm for shortest paths in road networks,'' in
  \emph{Proceedings of the 10th international conference on Experimental
  algorithms}, 2011.

\bibitem{Jensen:2003}
C.~S. Jensen, J.~Kol\'{a}\v{r}vr, T.~B. Pedersen, and I.~Timko, ``Nearest
  neighbor queries in road networks,'' in \emph{Proceedings of the 11th ACM
  international symposium on Advances in geographic information systems}, 2003,
  pp. 1--8.

\bibitem{Mouratidis:2006}
K.~Mouratidis, M.~L. Yiu, D.~Papadias, and N.~Mamoulis, ``Continuous nearest
  neighbor monitoring in road networks,'' in \emph{Proceedings of the 32nd
  international conference on Very large data bases}, ser. VLDB '06, 2006, pp.
  43--54.

\bibitem{Chen:2009:MPN}
Z.~Chen, H.~T. Shen, X.~Zhou, and J.~X. Yu, ``Monitoring path nearest neighbor
  in road networks,'' in \emph{Proceedings of the 2009 ACM SIGMOD International
  Conference on Management of data}, ser. SIGMOD '09, 2009, pp. 591--602.

\bibitem{Guting:2010}
R.~H. G\"{u}ting, T.~Behr, and J.~Xu, ``Efficient k-nearest neighbor search on
  moving object trajectories,'' \emph{The VLDB Journal}, vol.~19, no.~5, pp.
  687--714, Oct. 2010.

\bibitem{Psaraftis:1983}
H.~Psaraftis, ``An exact algorithm for the single-vehicle many-to-many
  dial-a-ride problem with time windows,'' \emph{Transportation Science},
  vol.~17, no.~3, pp. 351--357, 1983.

\bibitem{Coslovich:2006}
L.~Coslovich, R.~Pesenti, and W.~Ukovich, ``A two-phase insertion technique of
  unexpected customers for a dynamic dial-a-ride problem,'' \emph{European
  Journal of Operational Research}, vol. 175, no.~3, pp. 1605--1615, 2006.

\end{thebibliography}

\end{document}